\documentstyle[aps,twocolumn,amsmath,amssymb,psfig]{revtex}
\newcommand{\ie}{i.{\thinspace}e.}

\newcommand{\be}{\begin{equation}}
\newcommand{\ee}{\end{equation}}
\newcommand{\bea}{\begin{eqnarray}}
\newcommand{\eea}{\end{eqnarray}}
\newcommand{\ket}[1]{\left| #1\right \rangle}
\newcommand{\bra}[1]{\left \langle #1\right|}
\newcommand{\twoket}[2]{\left| #1\right \rangle
                        \otimes
                        \left| #2\right \rangle }
\newcommand{\twobra}[2]{\left \langle #1\right|
                        \otimes
                        \left \langle #2\right|}
\newcommand{\scal}[2]{\langle #1| #2 \rangle}
\newcommand{\av}[1]{\langle #1 \rangle}

\newcommand{\al}[1]{\alpha_{#1}^{\!\!\phantom{\ast}}}
\newcommand{\ald}[1]{\alpha_{#1}^\ast}

\newcommand{\alpp}[1]{ \alpha_{{#1}^\ppr}^{\!\!\phantom{\ast}}}

\newcommand{\piGP}[1]{\Pi_{{\cal{#1}}}}

\newcommand{\Gg}{G^{>}}
\newcommand{\Gk}{G^{<}}

\newcommand{\Upg}[1]{\Upsilon_{\cal{#1}}^{>}}
\newcommand{\Upk}[1]{\Upsilon_{\cal{#1}}^{<}}

\newcommand{\Gag}[1]{\Gamma_{\cal{#1}}^{>}}
\newcommand{\Gak}[1]{\Gamma_{\cal{#1}}^{<}}

\newcommand{\sigN}{\Sigma_{\cal{N}}}
\newcommand{\sigA}{\Sigma_{\cal{A}}}

\newcommand{\paul}[1]{\sigma_{{#1}}}

\newcommand{\aop}[1]{{ \hat{a}_{#1}^{\!\!\phantom{\dag}}}}
\newcommand{\aopd}[1]{{\hat{a}_{#1}^{\dag}}}

\newcommand{\Gcoll}[1]{{\Gamma_{#1}}}

\newcommand{\sign}[1]{{\mathbf{\sigma}}_{\{\gamma #1\}}^{(0)}}

\newcommand{\sigo}[1]{{\mathbf{\sigma}}_{\{\gamma #1\}}^{(1)}}

\newcommand{\bx}{{\mathbf{x}}}
\newcommand{\bp}{{\mathbf{p}}}
\newcommand{\by}{{\mathbf{y}}}

\newcommand{\hn}{\widehat{H}^{(0)}}
\newcommand{\ho}{\widehat{H}^{(1)}}

\newcommand{\ppr}{{\prime \prime}
}
\newcommand{\fs}[1]{\tilde{f}_{#1}}

\newcommand{\fspo}[1]{(1+\tilde{f})_{#1}}

\newcommand{\ms}[1]{\widetilde{m}_{#1}}

\newcommand{\ns}[1]{\widetilde{n}_{#1}}

\newcommand{\fc}[1]{f^{(c)}_{#1}}
\newcommand{\mc}[1]{m^{(c)}_{#1}}
\newcommand{\nc}[1]{n^{(c)}_{#1}}
\newcommand{\ft}[1]{f_{#1}}
\newcommand{\mt}[1]{m_{#1}}
\newcommand{\nt}[1]{n_{#1}}

\newcommand{\phiunpr}{\phi^{1 2 3 4}}
\newcommand{\phipr}{\phi^{1\,2^\prime 3^\prime 4^\prime}}
\newcommand{\phippr}[1]{\phi^{1^\ppr 2^\ppr 3^\ppr 4^\ppr}_{#1}}

\newcommand{\phiprb}{\phi^{\bar{1}\,\bar{2}^\prime \bar{3}^\prime 
\bar{4}^\prime}}
\newcommand{\phipprb}[1]{\phi^{\bar{1}^\ppr \bar{2}^\ppr \bar{3}^\ppr 
\bar{4}^\ppr}_{#1}}

\newcommand{\phireduc}[4]{\phi^{\bar{#1} \bar{#2} \bar{#3} \bar{#4}}}
\newcommand{\Yangul}[4]{Y^{{\underline{#1}}\, {\underline{#2}}\, 
{\underline{#3}}\, {\underline{#4}}}}
\newcommand{\gangul}[4]{g^{{\bar{#1}}\, {\bar{#2}}\, {\bar{#3}}\, {\bar{#4}}}}
\newcommand{\gangulprb}{g^{\bar{1} \bar{2}^\prime \bar{3}^\prime
\bar{4}^\prime}}

\newcommand{\eps}[1]{\varepsilon_{{#1}}}

\newcommand{\epspp}[1]{\varepsilon_{{#1}^{\prime \prime}}}

\newcommand{\Uc}{U_{\fc{}}}
\newcommand{\Usq}{U_{\fs{}}}

\newcommand{\Vsq}{V_{\ms{}}}
\newcommand{\Vm}{V_{(\mc{}+\ms{})}}

\newcommand{\BE}{Bose-{E}instein }

\begin{document}

\draft
\wideabs{ 

\title{Reversible and irreversible evolution of a
    condensed bosonic gas} \author{R.~Walser, J.~Cooper, and
    M.~Holland} \address{JILA, National Institute of Standards and
    Technology and University of Colorado, Boulder, CO 80309-0440}
  \date{August 25, 2000}
  \maketitle{}
  \begin{abstract}
    We have formulated a kinetic theory for a condensed atomic gas in
    a trap, \ie, a generalized Gross-Pitaevskii equation, as well as a
    quantum-Boltzmann equation for the normal and anomalous
    fluctuations \protect{[R. Walser {\em et al.}, Phys. Rev. A, {\bf 59}, 3878
      (1999)]}.  In this article, the theory is applied to the case of an
    isotropic configuration and we present numerical and analytical
    results for the reversible real-time propagation, as well as
    irreversible evolution towards equilibrium.
  \end{abstract}
 }


\pacs{PACS Nos. 03.75.Fi, 05.30.Jp, 67.40.Db, 05.70.Ln} 
\narrowtext
\section{Introduction}
More than 70 years ago, S. Bose and A. Einstein proposed a provocative
hypothesis --- that at ultra-low temperatures a novel state of matter
should exist. They predicted this state could be attained by cooling
an ordinary gas towards absolute zero.  At a well-defined point in
this process, a spontaneous transition should occur and change the
state of matter from an unordered ensemble of individual particles
into one collective entity. This single object, now devoid of its
many-particle character, ought to evolve as a collective matter wave.

With the discovery of superfluidity in liquid helium in 1938 and its
subsequent explanation in terms of \BE condensation (BEC), the
hypothesis had been firmly established.  In turn, this phenomenon has
had a major impact on the development of modern quantum physics.
Today, BEC is fundamental to our understanding of many low-temperature
phenomena and it is the cornerstone of many quantitative explanations.
However, up to 1995, condensation of a weakly interacting, atomic
\BE gas had never been achieved, as such.

With the ground-breaking accomplishment of condensing atomic
${}^{87}$Rb by E.~Cornell and C.~Wieman {\em et al.}
\cite{Anderson1995a}, of sodium by W.~Ketterle {\em et
  al.}\cite{Davis1995b}, and lithium by R. Hulet {\em et al.}
\cite{Bradley1997a}, a new chapter of quantum statistical physics has
been opened.  For the first time, it is possible to study in a
table-top experiment quantum statistical effects of material objects
on a human scale (up to 5 mm)---the very phenomena that govern the
otherwise microscopic physics of nuclear matter, macroscopic quantum
liquids, or astronomical objects, such as neutron stars.

Today, more bosonic alkali elements have crossed the transition
temperature, in particular atomic hydrogen \cite{Fried1998a} as well
as ${}^{85}$Rb \cite{cornish400}, and many more vastly improved
experiments have been carried out. For example, it is now possible to
examine multi-component condensates \cite{Esry1998b,Williams1999a}, to create
vortices \cite{holland999,cornell999}, and to prepare novel
topological modes \cite{williams2000a}.  For a list of current
experiments see Ref.\cite{webpage}, or the review article
in Ref.\cite{ketterleProc99}. However, the technological breakthrough of
combining laser cooling with evaporative cooling is not limited to
bosonic species only. Most recently, the fermionic isotope of
potassium ${}^{40}$K has also been cooled successfully below the
Fermi temperature \cite{jin999}.

Instigated by these spectacular experiments, strongly renewed interest
has developed in their quantitative description.  While cold quantum
gases had been studied extensively in the 1950-60s, they
were mainly considered as precursor theories for strongly interacting
systems, such as liquid helium. Thus, most of the available results
were focused on spatially uniform systems in thermal equilibrium. 
Excellent accounts of these standard results can be found, for example,
in the textbooks and monographs 
\cite{peletminskii,kadanoff62,blaizot,pines90,Griffin1995b,zubarev1}.
However, the spatial non-uniformity, the thermal isolation resulting
from the confinement in a ultra-high vacuum trap, as well as the large
disparity of collision and relaxation time-scales, are indispensable
ingredients for a quantitative description of today's experiments.

To account for these differences that distinguish the present
experimental situation from the homogeneous \BE gas
\cite{beliaev58a,kadanoff65,Hohenberg1965,Kirkpatrick1985b},
a growing number of equilibrium and nonequilibrium kinetic theories
have been recently presented
\cite{prouk298,zollerv,shlyapnikov1098,morgan99a,Stoof1999a,zaremba899}.
However, the effort to go beyond the mean-field description
of the Gross-Pitaevskii equation \cite{stringarireview} is
considerable.  Thus, the research for a unified description of the
equilibrium and nonequilibrium situation is still very active.

In this article, we explore numerically and analytically some of the
implications of the reversible and irreversible evolution of a
condensed gas immersed in the noncondensate cloud. The points
discussed are organized as follows. Section~\ref{kinetic} revisits the
main results of our kinetic theory \cite{Walser1999a}, \ie, the
two-particle Hamiltonian and the energy and number conserving
collisional kinetic equations for the condensate, as well as the
normal and anomalous fluctuations.  In Sec.~\ref{isotropic}, we
specialize these kinetic equations for a completely isotropic
situation.  Based on these prerequisites, we discuss in
Sec.~\ref{reversible} the results of propagating the collisionless
mean-field and the Hartree-Fock-Bogoliubov (HFB) equations in
real-time. Finally, in Sec.~\ref{irreversible}, we study the evolution
of an ergodic distribution towards equilibrium in the presence of
collisions.
\section{Kinetic master equations}
\label{kinetic}
\subsection{Master variables}
The kinetic master equation of the weakly interacting dilute atomic
gas describes the coupled evolution of the condensed fraction immersed
in the quantum fluctuations. In this context, we associate the
condensate with a c-number field $\al{\bx}(t)$ that represents the
expectation value of the quantum field $\av{\aop{\bx}(t)}$.  The field
operator $\aop{\bx}$ removes a particle from point $\bx$ and satisfies
the scalar, equal-time  commutation relation, \bea
[\aop{\bx},\aopd{\by}]&=&\delta(\bx-\by), \eea of a boson. The
position representation $\{\ket{\bx}\}$ used above is not necessarily
the most suitable basis to formulate a kinetic theory. It proves to be
more useful to postpone the choice of a particular representation and
to formulate the theory in terms of a general single particle basis
$\{\ket{i_1}\}$ that spans the same single-particle Hilbert space:
\bea \aop{\bx}=\sum_{i_1}\aop{i_1}\,\scal{\bx}{i_1}.  \eea

In the case of an unstructured (scalar) atomic condensate, three
external quantum labels ($i_1$) are sufficient to describe its
motional state in space, completely.\footnote{This is readily
  generalized to accommodate multiple internal electronic
  configurations if $i_1$ encompasses more quantum labels accordingly,
  \ie, $\ket{i_1}=\ket{n_1,l_1,m_1;F_1,M_1,\ldots}$.}
In this manner, we can expand any field as \bea
  \av{{\hat{a}}}&=&\sum_{i_1} \al{i_1}\,\ket{i_1} \equiv
  \al{1}\,\ket{1}\equiv \alpha.
  \eea 
  Here we have simplified the notation by dropping the name of the
  dummy variable, \ie, $i_1\equiv 1$, and by assuming implicit
  summation over repeated indices, as usual.

In an analogous fashion, we can describe the normal density of the
atomic gas $\ft{}=\av{ {\hat{a}}^{\dag} \hat{a} }=\fc{}+\fs{}$ by a Hermitian
tensor operator of rank (1,1):
\bea
\begin{array}{ll} 
\label{fcfsq}      
  \fs{}=\fs{12}\,\ket{1}\bra{2}, &
  \fc{}=\ald{2}\al{1}\,\ket{1}\bra{2}.
\end{array}
\eea 
Moreover, we will always decompose any quantum
average into  a mean-field contribution and the remaining fluctuations.
Similarly, we define the anomalous averages 
$\mt{}=\av{\hat{a} \hat{a}}=\mc{}+\ms{}$ 
as symmetric tensors of rank (2,0), 
\bea
\begin{array}{ll}
\label{mcmsq}       
  \ms{}=\ms{12}\,\ket{1}\ket{2}, & \mc{}=\al{2}\al{1}\,\ket{1}\ket{2},
\end{array}
\eea 
and  their symmetric conjugates as 
$\nt{}=\mt{12}^\ast\,\bra{1}\bra{2}$.
\subsection{Dynamical evolution}
The kinetic evolution of a weakly interacting gas is primarily
governed by the motion of the individual particles in the external
trapping potential and by binary collisions. Simultaneous collisions
of more than two particles are unlikely events in a dilute gas.
Consequently, we will disregard such processes and use the following
number-conserving Hamiltonian operator: \bea
\label{generic}
\widehat{H}=\hn+\ho&=&{H^{(0)}}^{12}\,\aopd{1} \aop{2}+
\phiunpr\,\aopd{1} \aopd{2}\aop{3}\aop{4}.  \eea 
Here, $\hn$ denotes a
single-particle Hamiltonian operator with matrix elements
${H^{(0)}}^{12}=\bra{1}\bp^2/(2\,m)+V_{{\text{ext}}}(\bx)\ket{2}$.
For the external trapping potential, we assume a three-dimensional
isotropic harmonic oscillator, 
$V_{\text{ext}}(\bx)=m\omega^2\,(x^2+y^2+z^2)/2$.  
In most of the present experiments with large, 
stable condensates, the two-body interaction potentials
$V_{\text{bin}}(\bx_1-\bx_2)$ are repulsive and of short range. From
such potentials, we can obtain two-particle matrix elements as
\bea
\label{2bdymatelem}
\phiunpr&=&\frac{1}{2}({\mathcal{S}})
\twobra{1}{2}V_{\text{bin}}(\bx_1-\bx_2)\twoket{3}{4},\\
\phiunpr&=&\phi^{1243}=\phi^{2134}=\phi^{2143}.  \eea Only the
symmetric part of the two-particle matrix element $\phiunpr$ is
physically relevant. Therefore, we have explicitly $(\mathcal{S})$
symmetrized it.  In the low kinetic energy range that we are
interested in, s-wave scattering is the dominant two-particle
scattering event
\cite{verhaar694,wieman495,julienne597}.  Thus,
by discarding all details of the two-particle potential, we can
describe the interaction strength with a single parameter $V_0$
related to the scattering length $a_{{\text{s}}}$ by $V_0=4\pi \hbar^2
a_{{\text{s}}}/m$. This limit corresponds to a singular interaction
potential, \ie, $V_{\text{bin}}(\bx_1,\bx_2)=V_0\,
\delta(\bx_1-\bx_2)$.  In the case of this delta potential, one finds
for the two-body matrix elements: \bea
\label{deltamatrixel}
\phiunpr&=& \frac{V_0}{2} \int d^3\bx
\scal{1}{\bx}\scal{2}{\bx}\scal{\bx}{3}\scal{\bx}{4}, \eea which need
not be symmetrized, as they are symmetric already.  However,
considering the caveats that are related to the singular functional
form of the two-particle potential \cite{huang}, we will only rely on
the existence and symmetry of the two-particle matrix elements as
defined in Eq.~({\ref{2bdymatelem}).

\subsection{Mean field equations}
Based on these assumptions, we have derived a set of kinetic equations
that describe the dynamical evolution of the condensate fraction
immersed in a cloud of noncondensate particles.  By discarding all of
the interactions except for the condensate's self-interaction, they
reduce to the familiar Gross-Pitaevskii (GP) equation for the mean
field $\alpha$.  However, due to the presence of anomalous
fluctuations $\ms{}$, this nonlinear, but otherwise unitary
GP equation acquires a contribution proportional to the
time-reversed or complex conjugated field $\ald{}$.

To represent these equations compactly, it is useful to arrange them
in a $2\times2$ matrix form.  Moreover, we transform this field
equation to a frame co-rotating with a positive frequency $\mu$
defined by $\alpha(t)=\exp{(-i \mu t)}\,{\overline{\alpha}}(t)$.
However, in order not to overload the notation, we will suppress the
overline in the following generalized GP-equation \bea
\label{GPeq}
\frac{d}{dt} \chi &=&\left(-i\,
  \piGP{}+\Upk{}-\Upg{}\right) \, \chi. \eea 
The two-component state vector $\chi=(\alpha,\ald{})^\top$, introduced above keeps
track of the forward and time-reversed components of the mean-field.
It is symmetric under time-reversal, \ie,
$\chi=\paul{1}\chi^\ast$. The  Pauli matrix $\paul{1}$ achieves the exchange
of upper and lower components and is defined in Appendix \ref{canonical}.

Two distinct processes govern the real-time evolution of the mean-field.
First, there is the generalized GP-propagator that is defined as
\bea
\piGP{}&=&
\left(
  \begin{array}{cc}
    \piGP{N}& \piGP{A}\\
    -\piGP{A}^\ast & -\piGP{N}^\ast
  \end{array}
\right). \eea  
The two contributions that define this symplectic propagator are 
a normal Hermitian Hamiltonian operator
\bea
\label{Hc}
\piGP{N}&=&H^{(0)}+1\,\Uc+2\,\Usq-\mu\,,
\eea
as well as a symmetric anomalous coupling strength
\bea
\label{Oc}
\piGP{A}&=&\Vsq.  \eea It is easy to identify $\piGP{N}$ with the well
known unitary GP-propagator that accounts for the free evolution of
the mean-field $(H^{(0)}-\mu)$, its self-interaction $\Uc$, as well as
the energy shift $\Usq$ caused by the presence of the noncondensate
cloud.  However, due to the existence of the anomalous fluctuations
there is also a coupling through $\piGP{A}$ to the time-reversed
field.  For convenience, we have introduced two auxiliary operators
$U_{\ft{}}$ and $V_{\mt{}}$.  Explicitly, they are defined in terms of
the two-body matrix elements, such as \bea U_{\ft{}}&=& 2\,\phipr
\,\ft{3^\prime 2^\prime}\,\ket{1}\bra{4^\prime}, \eea and a first
order anomalous coupling strength \bea V_{\mt{}}&=& 2\,\phipr
\,\mt{3^\prime4^\prime}\,\ket{1}\ket{2^\prime}. \eea Second, there are
all of the collisional second-order damping rates and energy
shifts\cite{beliaev58a,kadanoff65,Hohenberg1965,Kirkpatrick1985b} that
are given by \bea \Upk{}&=&\left(
\begin{array}{cc}
\Upk{N} & \Upk{A}\\
-{\Upg{A}}^\ast & -{\Upg{N}}^\ast
\end{array}
\right), \eea and the time-reversed contribution
$\Upg{}=-\paul{1}{\Upk{}}^\ast\paul{1}$. 
It can be shown that they are equivalent to the extended Beliaev
rates \cite{Wachter2000}.

The forward and backward transition rates $\Upk{N}$, $\Upk{A}$,
$\Upg{A}$, and $\Upg{N}$, describe the bosonically enhanced scattering
of noncondensate particles into and out of the condensate. In turn,
these transition rates are formed from various binary scattering
processes $\Gcoll{}$, and are given by \bea \Upk{N}&=& \Gcoll{\fs{}
  \fs{} \fspo{}}
+2\, \Gcoll{\fs{} \ms{} \ns{}},\\
\Upg{N}&=& \Gcoll{\fspo{} \fspo{} \fs{}} +2\,\Gcoll{\fspo{} \ms{}
  \ns{}}, \eea and \bea \Upk{A}&=& \Gcoll{\ms{} \ms{} \ns{}}
+2\,\Gcoll{\fs{} \ms{} \fspo{}},\\
\Upg{A}&=& \Gcoll{\ms{} \ms{} \ns{}} +2\,\Gcoll{\fspo{} \ms{}\fs{}}.
\eea Within the Born-Markov approximation of kinetic theory, we define
these elemental collision processes as \bea
\label{collops}
\Gcoll{\ft{} \ft{} \ft{}}&=& 8\,\phipr \phippr{\eta} \ft{3^\prime
  1^\ppr} \ft{4^\prime 2^\ppr}
\ft{4^\ppr 2^\prime}\,\ket{1} \bra{3^\ppr},\nonumber\\
\Gcoll{\ft{} \mt{} \ft{}}&=& 8\,\phipr \phippr{\eta} \ft{3^\prime
  1^\ppr} \mt{4^\prime 3^\ppr}
\ft{4^\ppr 2^\prime }\,\ket{1} \ket{2^\ppr},\nonumber\\
\Gcoll{\ft{} \mt{} \nt{}}&=& 8\,\phipr \phippr{\eta} \ft{3^\prime
  1^\ppr} \mt{4^\prime 3^\ppr}
\nt{2^\ppr 2^\prime}\,\ket{1} \bra{4^\ppr},\nonumber\\
\Gcoll{\mt{} \mt{} \nt{}}&=& 8\,\phipr \phippr{\eta} \mt{3^\prime
  4^\ppr} \mt{4^\prime 3^\ppr} \nt{2^\ppr 2^\prime}\,\ket{1}
\ket{1^\ppr}.\nonumber\\
\eea During a binary collision event, two particles can conserve their
energy only approximately. After all, the individual scattering event happens
within a medium and the asymptotics can not be reached within the finite duration 
of the collision. Thus, within the limits of the Born-Markov
approximation, any second order collision operator accrues a
dispersive as well as a dissipative part from the complex
valued matrix-element \bea \phippr{\eta}&=&\phippr{}\, \frac{1}{\eta-i\,
  \Delta_{1^\ppr2^\ppr 3^\ppr 4^\ppr}}.  \eea It is essentially
non-zero only if the energy difference $\Delta_{1^\ppr 2^\ppr 3^\ppr
  4^\ppr}=\varepsilon_{1^\ppr}(t)+
\varepsilon_{2^\ppr}(t)-\varepsilon_{3^\ppr}(t)-\varepsilon_{4^\ppr}(t)$
between the pre- and post-collision energies is smaller than an energy
uncertainty $\eta$: \bea \lim_{\eta\rightarrow 0_+}\frac{1}{\eta-i\,
  \Delta}&=&
\pi\,\delta_\eta(\Delta)+i\,{\mathcal{P}}_\eta\frac{1}{\Delta}. \eea
On general physical grounds, it can be argued that this uncertainty
$\eta$ is bracketed by the binary collision rate, on one side, and the
energy uncertainty arising from the finite duration of an individual
collision event on the other side.  As we have shown, one has also the
liberty to choose a more accurate intermediate propagator such that
the single particle energies $\varepsilon_{}(t)$ and the eigen-states
incorporate mean field shifts.
\subsection{Normal and anomalous fluctuations}
The normal and anomalous fluctuations $\fs{}(t)$ and $\ms{}(t)$ of a
quantum field are not independent quantities, but actually they are
the components a generalized single-time density operator $\Gg(t)$: \bea
\label{generaldensity}
\Gg&=&
\left(
  \begin{array}{cc}
    \fs{}& \ms{}\\
    \ns{}& \fspo{}^\ast
  \end{array}
\right)\ge 0.  \eea The non-negativity of this co-variance operator
implies that the magnitude of the anomalous fluctuations is limited by
the normal depletion through a Cauchy-Schwartz inequality (see
Appendix \ref{CauchySchwartz}).  In the general context of Green
function's \cite{kadanoff62,blaizot}, this single-time density
operator $\Gg(t)$ can also be viewed as a particular limit of a
time-ordered (${\cal T}$), two-time Green function $G(\tau,t)$,
\ie, $\Gg(t)=\lim_{\tau \rightarrow t_+} {\cal T}
\,G(\tau,t)$. Consequently, it is also necessary to consider the
opposite limit and to define a time-reversed, single-time density
operator through $\Gk(t)=\lim_{\tau\rightarrow t_-} {\cal T}\,
G(\tau,t)$.  Explicitly, this operator is given by \bea
\label{Gk}
\Gk&=&\paul{1}\, {\Gg}^\ast \paul{1}=\Gg+\paul{3}=
\left(
  \begin{array}{cc}
    1+\fs{}& \ms{}\\
    \ns{}& \fs{}^\ast
  \end{array}
\right).  \eea With the help of these definitions, we can
now present the results of the kinetic theory as a generalized
Boltzmann equation for the single-time density operator $\Gg(t)$ as
\bea
\label{Boltzeq}
\frac{d}{dt} \Gg &=&-i\,\Sigma\,\Gg + \Gak{}\,\Gk-\Gag{}\,\Gg
+\text{h.\thinspace c.} \eea 

In analogy to the previous discussion of the mean-field dynamics, we
again find that the evolution of the density operator is ruled by two
types of propagators.  First, there is the Hartree-Fock-Bogoliubov
(HFB) self-energy operator $\Sigma$ that can be obtained also by
variational methods \cite{blaizot}. In detail, this symplectic
self-energy is given by \bea \Sigma&=& \left(
  \begin{array}{cc}
    \sigN& \sigA\\
    -\sigA^{\ast} & -\sigN^\ast
  \end{array}
\right), \eea where we have introduced Hermitian Hamiltonian
operators and symmetric anomalous coupling potentials as \bea
\label{Hsq}
 \sigN&=&H^{(0)}+2\,\Uc+2\,\Usq-\mu,\\
\label{Osq}
\sigA&=&\Vm.  \eea It is important to note the different weighing
factors of the mean-field potential in Eqs.~(\ref{Hc}) and
(\ref{Hsq}), as well as the appearance of the anomalous condensate
density $\mc{}$ in Eq.~(\ref{Osq}).  This HFB-operator is the usual
starting point of any finite-temperature calculations. Depending on
additional considerations, \ie, `gapless vs. conserving
approximations' (see
Refs.~\cite{Hohenberg1965,griffin496,Lewenstein1996a,You1998,Castin1998a}),
the anomalous couplings $\Vsq{}$ are usually discarded from
Eqs.~(\ref{Oc}) and (\ref{Osq}).  However, since we do go beyond a
first order calculation, we need to retain all contributions for
consistency.
  
Second, the Boltzmann equation, Eq.~(\ref{Boltzeq}), introduces forward and
backward collision operators $\Gak{}$ and $\Gag{}$. They are
responsible for particle transfer out of and into the condensate on
one hand, and lead to thermal equilibration within the non-condensate
cloud, on the other hand.  These forward and backward collision
operator are defined by \bea \Gak{}&=&\left(
\begin{array}{cc}
\Gak{N} & \Gak{A}\\
-{\Gag{A}}^\ast & -{\Gag{N}}^\ast\\
\end{array}
\right),
\eea
and $\Gag{}=-\paul{1}{\Gak{}}^\ast\paul{1}$, where
\bea
\label{GakN}
\Gak{N}&=& \Gcoll{(\fs{}+\fc{}) \fs{} \fspo{}} +\Gcoll{\fs{} \fc{}
  \fspo{}}
+\Gcoll{\fs{} \fs{} \fc{}}\nonumber\\
&+& 2\,\left( \Gcoll{(\fs{}+\fc{}) \ms{} \ns{}} +\Gcoll{\fs{} \mc{}
    \ns{}} +\Gcoll{\fs{} \ms{} \nc{}}
\right),\\
\Gag{N}&=& \Gcoll{(1+\fs{}+\fc{}) \fspo{} \fs{}} +\Gcoll{\fspo{} \fc{}
  \fs{}}
+\Gcoll{\fspo{} \fspo{} \fc{}}\nonumber\\
&+& 2\,\left( \Gcoll{(1+\fs{}+\fc{}) \ms{} \ns{}} +\Gcoll{\fspo{}
    \mc{} \ns{}} +\Gcoll{\fspo{} \ms{} \nc{}} \right), \eea and \bea
\Gak{A}&=& \Gcoll{(\ms{}+\mc{}) \ms{} \ns{}} +\Gcoll{\ms{} \mc{}
  \ns{}}
+\Gcoll{\ms{} \ms{} \nc{}}\nonumber\\
&+& 2\,\left( \Gcoll{(\fs{}+\fc{}) \ms{} \fspo{}} +\Gcoll{\fs{} \mc{}
    \fspo{}} +\Gcoll{\fs{} \ms{} \fc{}}
\right),\\
\Gag{A}&=& \Gcoll{(\ms{}+\mc{}) \ms{} \ns{}} +\Gcoll{\ms{} \mc{}
  \ns{}}
+\Gcoll{\ms{} \ms{} \nc{}}\nonumber\\
&+& 2\,\left( \Gcoll{(1+\fs{}+\fc{}) \ms{} \fs{}} +\Gcoll{\fspo{}
    \mc{} \fs{}} +\Gcoll{\fspo{} \ms{} \fc{}} \right).  \eea It is
interesting to note that all of the collision processes that
contribute to the Boltzmann equation, Eq.~(\ref{Boltzeq}), are of the
same basic structure as the collision operators in the GP equation,
Eq.~(\ref{GPeq}).  In particular, one can generate all of the
processes $\Gak{}$ and $\Gag{}$ by functional differentiation from
$\Upk{}$ and $\Upg{}$ .  This very fact is actually is the key
principle to the functional-analytic Green functions method described
in Ref.~\cite{kadanoff62} and, for example, leads to the gapless
Beliaev approximation \cite{Hohenberg1965,Griffin2000}.

\subsection{Conservation laws}
\subsubsection{Number}
The total particle number $\hat{N}$ is a conserved quantity if the
atoms evolve under the generic two-particle Hamiltonian operator
$\hat{H}$ given by Eq.~(\ref{generic}), \ie,
$[\hat{H},\hat{N}]=0$.  This conservation law implies that the system
is invariant under a global phase change $\aop{}\rightarrow
\aop{}\,\exp{(i\,\varphi)}$.  By using this continuous symmetry,
\ie, $\al{}\rightarrow \al{}\,\exp{(i\,\varphi)}$,
$\fs{}\rightarrow \fs{}$, and $\ms{}\rightarrow \ms{}\,
\exp{(2\,i\,\varphi)}$, it is easy to see that the kinetic equations
Eqs.~(\ref{GPeq}) and (\ref{Boltzeq}) are also explicitly number
conserving at all times: \bea
\av{\hat{N}(t)}&=&\text{Tr}\{\fc{}(t)\}+\text{Tr}\{\fs{}(t)\}=\text{const.}
\eea Nevertheless, it is important to note that there are always
coherent and incoherent processes present that do transfer particles between
the condensate and the noncondensate clouds, continuously.

\subsubsection{Energy}
In the absence of any time-dependent external driving fields, such as
optical lasers or magnetic rf-fields, the overall energy $\hat{H}$
must be conserved as well.  To find the expectation value of the
total system energy  $E=\av{\hat{H}}=\text{Tr}\{\hat{H}\sigma(t)\}$, one
can use the same power-series expansion of the coarse-grained many-particle
density matrix $\sigma(t)$ that leads to the kinetic equations. Thus,
within the limits of
the Born-Markov approximation and the systematic application of Wick's
theorem, we have obtained first and second order contributions for the
energy 
$E=\text{Tr}\{\hat{H}\,(\sign{(t)}+\sigo{(t)})\}+{{\cal O}[3]}$. 
Explicitly, this energy functional $E=E^{(c)}+E[\fs{}]+E[\ms{}]$, is given as
\bea
\label{Ec}
E^{(c)}&=&\text{Tr} \{ 
  [
    H^{(0)}+\frac{1}{2}(1\,\Uc+2\,\Usq +i\,(\Upk{N}-\Upg{N}) )
  ]\, \fc{}\nonumber\\
&+&\frac{1}{2}\, (\Vsq{}+i\,(\Upk{A}-\Upg{A}))\,\nc{}\},\\
\label{Efs}
E[\fs{}]&=&\text{Tr}\{ 
  [ H^{(0)}+\frac{1}{2}\,(2\,\Uc+2\,\Usq
  +i\,(\Gak{N}-\Gag{N}))
  ] \,\fs{}\},\nonumber\\
\\
\label{Ems}
E[\ms{}]&=&\text{Tr} \{
\frac{1}{2}\,(V_{\mc{}+\ms{}}+i\,(\Gak{A}-\Gag{A})) \,\ns{}\}.  \eea
For example, the same first order results can be found in
Ref.\cite{blaizot}, derived by a variational procedure.
\section{A completely isotropic system}
\label{isotropic}
In the previous section, we have reviewed the main results of the
kinetic theory that describes the coupled evolution of the condensate
immersed in the non-condensate. The formal derivation did not rely on
a particular trapping geometry, nor a special form for the binary
interaction potential. In order to gain deeper understanding of the
intricate interactions, we will now specialize the theory to the most
simple, though realistic, three-dimensional model: a completely isotropic
configuration, a spherically symmetric harmonic trapping potential, and
a binary short-range s-wave scattering potential. 

\subsection{Irreducible tensor fields} 
Complete isotropy is easily achieved for the
mean-field by decomposing it in terms of a few zero angular momentum
partial waves.  For this purpose, we will use a set of basis states
$\{\ket{1}\}$ that can be characterized by radial and angular
momentum quantum numbers $(n,l,m)$, \ie,
$\scal{\bx}{1}=R^{n_1}_{l_1}(r)\,Y^{l_1}_{m_1}(\theta,\varphi).$

However, in order to isolate the isotropic components of the
non-condensate fluctuations, $\fs{}$ and $\ms{}$, we need to
generalize the concept of partial waves and introduce an irreducible
set of tensor fields.  Furthermore, by only selecting the scalar
component ($l=0$), we can enforce the desired radial symmetry.  Thus,
according to Refs.\cite{omont77,biedenharn}, we introduce irreducible
representations of tensor fields of rank: (2,0), (1,1), (0,2) as \bea
\label{Tlm}
T^{l}_{m}(\bar{1}\, \bar{2})&=& \sum_{m_1,m_2} (-1)^{l_2-m_2} C^{l_1
  l_2 l}_{m_1 -m_2 m}
\ket{n_1 l_1 m_1}\bra{n_2 l_2 m_2},\nonumber\\
S^{l}_{m}(\bar{1}\, \bar{2})&=& \sum_{m_1,m_2} (-1)^{l_2} C^{l_1 l_2
  l}_{m_1 m_2 m}
\ket{n_1 l_1 m_1 }\ket{n_2 l_2 m_2}.\nonumber\\
\eea The quantum labels carry additional overlines or underlines to
indicate whether a function depends only on two of the three quantum
labels, for example, $\overline{1}\equiv (n_1,l_1)$ or
$\underline{2}\equiv (l_2, m_2)$.  With these definitions, it is easy
to verify the following orthogonality relationships: \bea
\text{Tr}\left\{ T^{l}_{m}(\bar{1}\, \bar{2})\,
  {T^{l^\prime}_{m^\prime}(\bar{3}\,\bar{4})}^\dag
\right\}&=&\delta_{\bar{1}\bar{3}}\,\delta_{\bar{2}\bar{4}}\,
\delta_{l l^\prime} \,\delta_{ m m^\prime},\\
\text{Tr}\left\{ S^{l}_{m}(\bar{1}\, \bar{2})
  \,{S^{l^\prime}_{m^\prime}(\bar{3}\,\bar{4})}^\dag
\right\}&=&\delta_{\bar{1}\bar{3}}\,\delta_{\bar{2}\bar{4}}\,
\delta_{l l^\prime}\, \delta_{m m^\prime}.  \eea In the case of a
scalar field ($l=0$), one can simplify Eq.~(\ref{Tlm}) by the
following relation for the Clebsch-Gordan coefficients $C^{l_1 l_2
  0}_{m_1 m_2 0}=\delta_{l_1 l_2}\delta_{m_1, -m_2}
(-1)^{l_1-m_1}/\sqrt{2 l_1+1}$.  Provided the basis states transform
under a coordinate rotation, here denoted by $\mathcal{R}$, according
to the finite dimensional representation of the rotation group
${\mathcal{D}}^{l}({\mathcal{R}})$ \cite{biedenharn}, \bea
{\mathcal{U_R}} \ket{n l m}&=& \sum_{m^\prime}\ket{n l
  m^\prime}\,{\mathcal{D}}^{l}_{m^\prime m}({\mathcal{R}})\,; \eea it
follows that the set of tensors $\{T^{l}_{m}|\!|\, |m|\leq l\}$ and
$\{S^{l}_{m}|\!|\, |m|\leq l\}$ are irreducible as well: \bea
{\mathcal{U_R}}\otimes {\mathcal{U_R}}^{-1}
\,T^{l}_{m}(\bar{1}\,\bar{2})&=&
\sum_{m^\prime}T^{l}_{m\prime}(\bar{1}\,\bar{2})
\,{\mathcal{D}}^{l}_{m^\prime m}({\mathcal{R}}),\\
{\mathcal{U_R}}\otimes {\mathcal{U_R}}
\,S^{l}_{m}(\bar{1}\,\bar{2})&=&
\sum_{m^\prime}S^{l}_{m\prime}(\bar{1}\,\bar{2})
\,{\mathcal{D}}^{l}_{m^\prime m}({\mathcal{R}}). \eea

\subsection{Isotropic two-particle matrix element}
The most commonly used model for a short-range binary interaction
potential is the s-wave hard-core delta potential
$V_{\text{bin}}(\bx_1,\bx_2)=V_0\,\delta(\bx_1-\bx_2)$.  This model
potential is most suited to describe the low energetic collision
dynamics of two real particles.  However, it has to be used with
caution in connection with infinite summation over virtual, high
energy states.  It is clear that the energy independent scattering
approximation fails above a certain energy range when the spatial
scale of variation of the high-energetic wave functions begin to
sample the detailed form of the interaction potential.  Thus, the true
value of the interaction matrix element ought to decrease much faster
with energy than the value obtained from the simple hard core
delta-potential approximation.  It is well known that indiscriminate
use of the energy independent approximation leads to a non physical
ultra-violet divergence \cite{Stoof1999a,huang,Hutchinson1998a}.

Considering these limitations, we will use the s-wave scattering
matrix element that is obtained from the energy independent
approximation, \ie,  \bea \phiunpr&=& \frac{V_0}{2}
\int_{-\infty}^{\infty}d^3\bx\,
\scal{1}{\bx}\scal{2}{\bx}\scal{\bx}{3}\scal{\bx}{4},  \eea only for
energies below a certain level and truncate it appropriately
otherwise.

By using the basis states 
$ \scal{\bx}{1}=R^{n_1}_{l_1}(r)\,Y^{l_1}_{m_1}(\theta,\varphi),$
we can decompose the matrix element $\phiunpr$ into a reduced radial part 
$\phireduc{1}{2}{3}{4}$ and a purely 
geometric factor  $\Yangul{1}{2}{3}{4}$,
\bea
\label{swavematel}
\phiunpr&=&\phireduc{1}{2}{3}{4}\,\Yangul{1}{2}{3}{4},
\eea
where
\bea
\phireduc{1}{2}{3}{4}&=&
\frac{V_0}{2}\int_{0}^{\infty} r^2 dr \,
R^{n_1}_{l_1}(r) R^{n_2}_{l_2}(r) R^{n_3}_{l_3}(r) R^{n_4}_{l_4}(r)
\eea
and 
\bea
\label{Yangul}
\lefteqn{\Yangul{1}{2}{3}{4}=\int d^2{{\bf \Omega}}\,
{Y^{l_1}_{m_1}}^{\!\!\ast}({\bf \Omega})\,
{Y^{l_2}_{m_2}}^{\!\!\ast}({\bf \Omega})\,
Y^{l_3}_{m_3}({\bf \Omega})\,
Y^{l_4}_{m_4}({\bf \Omega})}\nonumber\\
&=&
\sum_{l=0}^{\infty}\frac{C^{l_1 l_2 l}_{0 0 0}
C^{l_3  l_4 l}_{0 0 0} C^{l_1  l_2 l}_{m_1 m_2 (m_1+m_2)}
C^{l_3  l_4 l}_{m_3 m_4 (m_1+m_2)}}{
4\pi\, (2 \,l+1) \,
[\prod_{i=1}^{4}(2\, l_i+1)]^{-1/2}}.
\eea 

In the context of evaluating the collision integrals, it will be
necessary to consider products of two matrix elements summed over all
energetically accessible sublevels. Within the isotropic model, all
magnetic sublevels are energetically degenerate. Moreover,
spherical symmetry demands an equal population distribution and
rules out the existence of coherences within magnetic sub-manifolds.
Thus, it will be required to know the magnitude of $|\Yangul{1}{2}{3}{4}|^2$ 
averaged over all the magnetic quantum numbers. For later reference,
we will now introduce such conveniently scaled factors as
\bea
g^{\overline{1}\,\overline{2}}&=&
((2\,l_1+1)(2\,l_{2}+1))^{1/2}/{2\pi},\\
\label{gangul}
\gangul{1}{2}{3}{4}&=& 2\,(\pi^2\,g^{\overline{1}\,\overline{2}}
g^{\overline{3}\,\overline{4}})^{-1}\,
\sum_{m_1 m_2 m_3 m_4} |\Yangul{1}{2}{3}{4}|^2 \nonumber\\
&=&2\,g^{\overline{1}\,\overline{2}}\, g^{\overline{3}\,\overline{4}}
\sum_{l=0}^{\infty}\frac{ (C^{l_1 l_2 l}_{0 0 0} C^{l_3 l_4 l}_{0 0
    0})^2}{(2 \,l+1)}.  \eea These coupling strengths
$\gangul{1}{2}{3}{4}$ in Eq.~(\ref{gangul}) measure the amount and
principle connectivity between pre-collision and post-collision
angular momenta sub-manifolds $(l_3,l_4)\rightarrow (l_1,l_2)$.  In
particular, it establishes a parity selection rule such that the
coefficients are non-vanishing only if the sum of the angular momenta
is $l_1+l_2+l_3+l_4= \mbox{even}$. In addition, transitions are
allowed only if the angular momentum $l$ is within a range of
$\mbox{max}(|l_1-l_2|,|l_3-l_4|)\leq l \leq
\mbox{min}(l_1+l_2,l_3+l_4)$.

\subsection{Scalar component of states and energies}
With the help of the auxiliary results established in the previous
section, we are now able perform the desired multi-pole decomposition
of the kinetic equations. The postulate of complete isotropy then
implies that we can focus on the scalar component of the field ($l=0$)
exclusively. This is 
\bea
\begin{array}{cc}
  \al{}=\delta_{l,0}\,\delta_{m,0}\,\al{1}\ket{1},&
\fs{}=\fs{\bar{1}\bar{2}}\,T^{0}_{0}(\bar{1}\,\bar{2}),\\
\ms{}=\ms{\bar{1}\bar{2}}\,S^{0}_{0}(\bar{1}\,\bar{2}),&
\ns{}=\ns{\bar{1}\bar{2}}\,{S^{0}_{0}}^\dag(\bar{1}\,\bar{2}).  
\end{array}
\eea Analogously, we can decompose all normal operators, such as the
bare single particle Hamiltonian operator, as 
\bea
H^{(0)}&=&{H^{(0)}}_{\bar{1}\,\bar{4}^\prime}\,
T^{0}_{0}(\bar{1}\,\bar{4}^\prime). 
\eea
The first order normal mean-field potential $U_{\ft{}}$
and the anomalous coupling strength $V_{\mt{}}$ 
are then
\bea
U_{\ft{}}&=&\phiprb\,g^{\overline{1}\,\overline{3}^\prime}
\delta_{l_{2^\prime} l_{3^\prime}}\, 
\ft{\bar{3}^\prime
  \bar{2}^\prime} \, T^{0}_{0}(\bar{1}\,\bar{4}^\prime),\\
V_{\mt{}}&=& 
\phiprb\, g^{\overline{1}\,\overline{3}^\prime}\,
\delta_{l_{3^\prime} l_{4^\prime}}\, 
\mt{\bar{3}^\prime \bar{4}^\prime} \,
\,S^0_0(\bar{1}\,\bar{2}^\prime).
\eea 
Finally, the normal and anomalous collisional contributions simplify to
\bea 
\Gcoll{\ft{} \ft{}\ft{}}&=&
\phiprb \phipprb{\eta}\,\gangulprb \,
\delta_{l_{3^\prime} l_{1^\ppr}}\,
\delta_{l_{4^\prime} l_{2^\ppr}}\,
\delta_{l_{4^\ppr} l_{2^\prime}} \nonumber\\
&\times& 
\ft{\bar{3}^\prime \bar{1}^\ppr} \, 
\ft{\bar{4}^\prime \bar{2}^\ppr} \, 
\ft{\bar{4}^\ppr \bar{2}^\prime} \, 
T^{0}_{0}(\bar{1}\,\bar{3}^{\prime \prime}),\\
\Gcoll{\ft{} \mt{} \nt{}}&=&
\phiprb \phipprb{\eta}\,\gangulprb  
\delta_{l_{3^\prime} l_{1^\ppr}}
\delta_{l_{4^\prime} l_{3^\ppr}}
\delta_{l_{2^\ppr} l_{2^\prime}}
\nonumber\\
&\times& \ft{\bar{3}^\prime \bar{1}^\ppr}\, 
\mt{\bar{4}^\prime \bar{3}^\ppr}\, 
\nt{\bar{2}^\ppr \bar{2}^\prime}\,
T^{0}_{0}(\bar{1}\,\bar{4}^\ppr),\\
\Gcoll{\ft{} \mt{} \ft{}}&=& 
\phiprb \phipprb{\eta}\,\gangulprb 
\delta_{l_{3^\prime} l_{1^\ppr}}\,
\delta_{l_{4^\prime} l_{3^\ppr}}\,
\delta_{l_{4^\ppr} l_{2^\prime}}\nonumber\\
&\times&
\ft{\bar{3}^\prime\bar{1}^\ppr}\, 
\mt{\bar{4}^\prime \bar{3}^\ppr}\, 
\ft{\bar{4}^\ppr \bar{2}^\prime}\,
S^{0}_{0}(\bar{1}\,\bar{2}^\ppr),\\
\Gcoll{\mt{}\mt{} \nt{}}&=& 
\phiprb \phipprb{\eta}\,\gangulprb 
\delta_{l_{3^\prime} l_{4^\ppr}}\,
\delta_{l_{4^\prime} l_{3^\ppr}}\,
\delta_{l_{2^\ppr} l_{2^\prime}}\,
\nonumber\\
&\times& 
\mt{\bar{3}^\prime \bar{4}^\ppr}\, 
\mt{\bar{4}^\prime \bar{3}^\ppr}\, 
\nt{\bar{2}^\ppr \bar{2}^\prime}\,
S^{0}_{0}(\bar{1}\,\bar{1}^\ppr).\eea
\section{Reversible evolution}
\label{reversible}
In this section, we will examine several limiting situations of the
reversible evolution in order to elucidate the complex behavior of the
condensed gas.  Since canonical transforms and Hartree-Fock-Bogoliubov
(HFB) operators are crucial for an understanding of the reversible
evolution, we will review the main results \cite{blaizot}.
Subsequently, we are going to examine the stationary equilibrium, as
well as the reversible real-time evolution of the condensed gas.

\subsection{Structure of the generalized density matrix}
\label{secG}
The definition of a generalized density matrix $G$, \ie,
either $\Gg$ or $\Gk$, was given in Eqs.~(\ref{generaldensity}) and
(\ref{Gk}). Its specific structure implies various important physical
properties.

First of all, we have to assume that there is a basis that
diagonalizes this $(2\,n\times 2\,n)$-dimensional fluctuation matrix.
Exactly $n$ of its $2\,n$ eigen-values correspond to the positive
occupation numbers of finding a particle or, more generally, a
quasi-particle in a certain mode.  For a given, but otherwise
arbitrary, $G$ matrix, one can construct this basis by studying the
transformation law of the density matrix under a canonical
transformation $T$ (see Appendix \ref{canonical}), \bea G^\prime=T\,
G\, T^\dag. \eea It is important to note that this is not the
transformation law of a general matrix under coordinate change. This
would require that $T^\dag=T^{-1}$.  However, by only using the
properties of the symplectic transformations, one can show that a
canonical eigen-value problem is defined by \bea \left( \paul{3}\,G
\right) \,T^\dag &=&T^\dag \, \left( \paul{3}\, G^\prime \right). \eea
The solution of this eigen-value problem yields the eigen-vector
matrix $T^\dag$ and the corresponding diagonal eigen-value matrix
$\paul{3}\,G^\prime$.  Here, we have introduced standard Pauli spin
matrices $\paul{1}$ and $\paul{3}$, which exchange upper and lower
component of a $2\,n$ dimensional vector, or flip the sign of the
lower segment, respectively.  All normalizable states can be rescaled
such that $T\paul{3}\,T^\dag=\paul{3}$.  Now, we are able to
reconstruct the positive $G$ matrix \bea \label{specdecG}
G&=&V\,P\,V^\dag,\eea from its eigen-vectors $V=\paul{3}\,T^\dag$ and
the diagonal, positive occupation number matrix $P=\paul{3}\,G^\prime
\paul{3}$.

Second, an important feature of an admissible fluctuation matrix is
its consistency with the commutation relation, \ie,
$\av{\aop{1}\aopd{2}}=\av{\aopd{2}\aop{1}}+\delta_{12}$ and
$\av{\aop{1}\aop{2}}=\av{\aop{2}\aop{1}}$.  This can be expressed
compactly as \bea \paul{1}\,G^\ast\,\paul{1} - G &=& \paul{3}.  \eea
By invoking the properties of a unitary symplectic transformation, one
can show that the elements of the diagonal occupation number matrix 
 $P$ are not $2\,n$ independent
variables. Actually half of them are determined by the other half,
$P_{(n+1,\dots, 2\,n)}=1+P_{(1,\dots,n)}$, or \bea \paul{1} \, P\,
\paul{1}-P=\paul{3}.  \eea In other words, by separating the
occupation numbers $P$ and the eigen-vector matrix $V$ into a first
and second half, \ie  $(P_{+},1+P_{+})=P$ and
$(V_{+},V_{-})=V$, one can then decompose a general fluctuation matrix
as \bea
\label{genericgmat}
G&=V_{+}^{\phantom{\dag}}\, P_{+}^{\phantom{\dag}} \, V_{+}^\dag 
+V_{-}^{\phantom{\dag}}\,(1+P_{+}^{\phantom{\dag}})\,V_{-}^\dag. \eea
\subsection{Structure of the Hartree-Fock-Bogoliubov operator}
\label{secHFB}
The symplectic HFB operator arises not only naturally in kinetic
theories or variational calculations, but in many other contexts
involving stability analysis.  In the case of bosonic fields, the
self-energy operator is of the generic form: \bea \Sigma&=& \left(
  \begin{array}{cc}
    \sigN & \sigA\\
    -{\sigA}^\ast & -{\sigN}^\ast
  \end{array} 
\right).  \eea In here, $\sigN$ stands for a Hermitian operator
$\sigN=\sigN^\dag$ and $\sigA$ denotes an anomalous coupling term that
has to be symmetric $\sigA=\sigA^\top$.  The relative size of the
operators $\sigN$ and $\sigA$ determines the character of the energy
spectrum. It can either be real valued with pairs of positive and
negative eigen-energies, or one finds a doubly degenerate zero
eigen-value, if the energy difference between the smallest positive
and highest negative vanishes (gapless spectrum).  In the general
case, there is a mixed spectrum consisting of pairs of real
sign-reversed as well as pairs of complex conjugated eigen-values. The
eigen-vectors $W$ are normalizable with respect to the indefinite norm
$|\!|W|\!|^2=W^\dag\paul{3}W$, except for those that belong to zero or
complex eigen-values.  It is important to note that this energy basis $W$ is
in general 
distinct from the instantaneous basis $V$ that diagonalizes the fluctuation
matrix $G$ in Eq.~(\ref{specdecG}). They do coincide only in equilibrium.
The mathematical properties of the eigen-states $W$ can be derived
easily from the intrinsic symmetries of the HFB operator: \bea
\label{sym1}
\Sigma&=&-\paul{1}\,\Sigma^\ast \paul{1},\\
\label{sym2}
\Sigma^\dag&=&\paul{3}\,\Sigma \,\paul{3}. 
\eea
Thus, if $W$
and $E$ are the solutions of the right eigen-value problem, 
\bea 
\label{rightev}
\Sigma\,W=W\,E, \eea 
it follows directly from Eq.~(\ref{sym1}) that
$\overline{W}=\paul{1}\,W^\ast$, is also a right eigen-vector but
corresponds to the eigen-value $\bar{E}=-E^\ast$. 
Starting from the second symmetry in Eq.~(\ref{sym2}) and the right
eigen-value problem of Eq.~(\ref{rightev}), it is easy to construct
the left eigen-vectors $\widetilde{W}=W^\dag\, \paul{3}$ that correspond
to the eigen-values $\tilde{E}=E^\ast$: \bea \widetilde{W}\,
\Sigma=E^\ast\,\widetilde{W}.  \eea
Finally, from a combination of the results for the right and left
eigen-vectors, it follows that the eigen-vectors are orthogonal with
respect to the metric $\paul{3}$: \bea
0=(E^\ast-E^\prime)\,W_{E}^\dag\paul{3}
W_{{E^\prime}}^{\phantom{\dag}}, \eea if $E^\ast \neq E^\prime$. On
the other hand, this relation implies also that eigen-vectors that
belong to complex eigen-values must have zero norm.

The situation of a doubly degenerate zero energy eigen-value $E=0$
needs special attention.  One can view this case as a limit when two
non-degenerate states approach each other. However, as the energy gap
decreases, the two eigen-states become more and more collinear. Thus,
in the limit of a vanishing energy separation, the dimension of the
spanned vector space collapses from 2 to 1 and $\Sigma$ becomes
defective. In the present context however, we did not encounter this
situation (see Ref.~\cite{blaizot} for details).
\subsection{Stationary solution of the Hartree-Fock-Bogoliubov equations}
In spite of the complex nonlinear interactions taking place within the
atomic gas, the kinetic evolution is completely reversible if we
disregard all collisionally induced redistributions of
quasi-particles.  Thus, for the moment, we will eliminate the collision
operators $\Upsilon$ and $\Gamma$ from the kinetic Eqs.~(\ref{GPeq})
and (\ref{Boltzeq}) and we will study the collisionless stationary
equilibrium, as well as the real-time evolution in this section.

With these assumptions, we are left with the following set of
stationary equations for the mean field $\chi$ and the
fluctuations $\Gg$:  
\bea
\label{firstorderequilibriumc}
0&=&\Pi\, \chi,\\
\label{firstorderequilibriumsq}
0&=&\Sigma\,\Gg-\Gg\,\Sigma^\dag.  \eea The self-energies of the
condensate $\Pi$ and the noncondensate $\Sigma$ are nonlinearly
coupled and implicitly include the rotation frequency of the
mean-field $\mu$.
 
However, the equilibrium solution to
Eqs.~(\ref{firstorderequilibriumc}) and
(\ref{firstorderequilibriumsq}) is not fully determined as it stands.
From the results of the previous section, we know that any fluctuation
matrix $\Gg$ that is diagonal with respect to the positive and negative
energy eigen-vectors $W=(W_+,W_-)$ of $\Sigma$, will be a stationary
and complete solution of Eq.~(\ref{firstorderequilibriumsq})
\bea
\label{statgmat}
\Gg&=W_{+}^{\phantom{\dag}}\, P_{+}^{\phantom{\dag}} \, W_{+}^\dag
+W_{-}^{\phantom{\dag}}\,(1+P_{+}^{\phantom{\dag}})\,W_{-}^\dag.\eea
By choosing a canonical Bose-Einstein distribution
$P(E>E_0)=1/[\exp{(\beta E)}-1]$ for the quasi-particles above the
non-degenerate ground state $E_0>0$ and a vanishing ground state
occupation number $P(E_0)=0$, we obtain a variationally minimal energy
solution for the total system at some inverse temperature $\beta$
\cite{blaizot}: \bea
\label{finiteTG}
\Gg&=&\sum_{E\ge E_0} P(E)\, W_{E}^{\phantom{\dag}} W_{E}^\dag+ 
\left(1+P(E)\right)\, W_{-E}^{\phantom{\dag}} W_{-E}^\dag. \eea

In order to understand the self-consistent equilibrium solution of
Eqs.~(\ref{firstorderequilibriumc}) and
(\ref{firstorderequilibriumsq}), it is useful to examine first the
potentials that govern the evolution of the condensate, as well as the
noncondensate.  In Fig.~\ref{fig_pot}, we depict the potential energy
densities of the normal Hamiltonian operators $\piGP{N}$ and $\sigN$
versus radius that arise for the zero angular momentum manifold $l=0$,
\ie, $V_{\text{ext}}+1\,\Uc$ and $V_{\text{ext}}+2\,\Uc$,
respectively.  They are compared to the bare isotropic harmonic
oscillator potential $V_{\text{ext}}=r^2/2$, for reference. Here and
in all of the subsequent results, we will use the experimental data of
a typical $^{87}$Rb condensate \cite{Jin1997a}.  All physical
parameters are scaled in the natural units for a harmonic oscillator,
\ie, the angular frequency $\omega=2\pi\,200\,\text{Hz}$,
the atomic mass $m_{87}=86.9092\, \text{amu}$, the ground state size
$a_{H}=[\hbar/(\omega \, m_{87})]^{1/2}= 763\,\text{nm}$, the s-wave
scattering length $a_{S}=5.82\,\text{nm}=7.63\,10^{-3} \,a_H$, a very
low temperature of $k_B T=0.2 \, \hbar\omega$, and a condensate number
chosen as $N^{(c)}=10^4$.  The isotropic particle densities are
normalized to $N=\int_{0}^{\infty} dr\,r^2\,f(r)$. From the effective
coupling parameter $\kappa=a_S/a_H$, one obtains an estimate of the
mean-field energy shift as $\mu_{\text{TF}}=(15
N^{(c)}\kappa)^{2/5}/2=8.36$. This gives an excellent approximation of
the self-consistent chemical potential of $\mu=8.52$, as can be seen in
Fig.~\ref{fig_pot}.
\begin{figure}[f]
  \begin{center}
    \psfig{figure=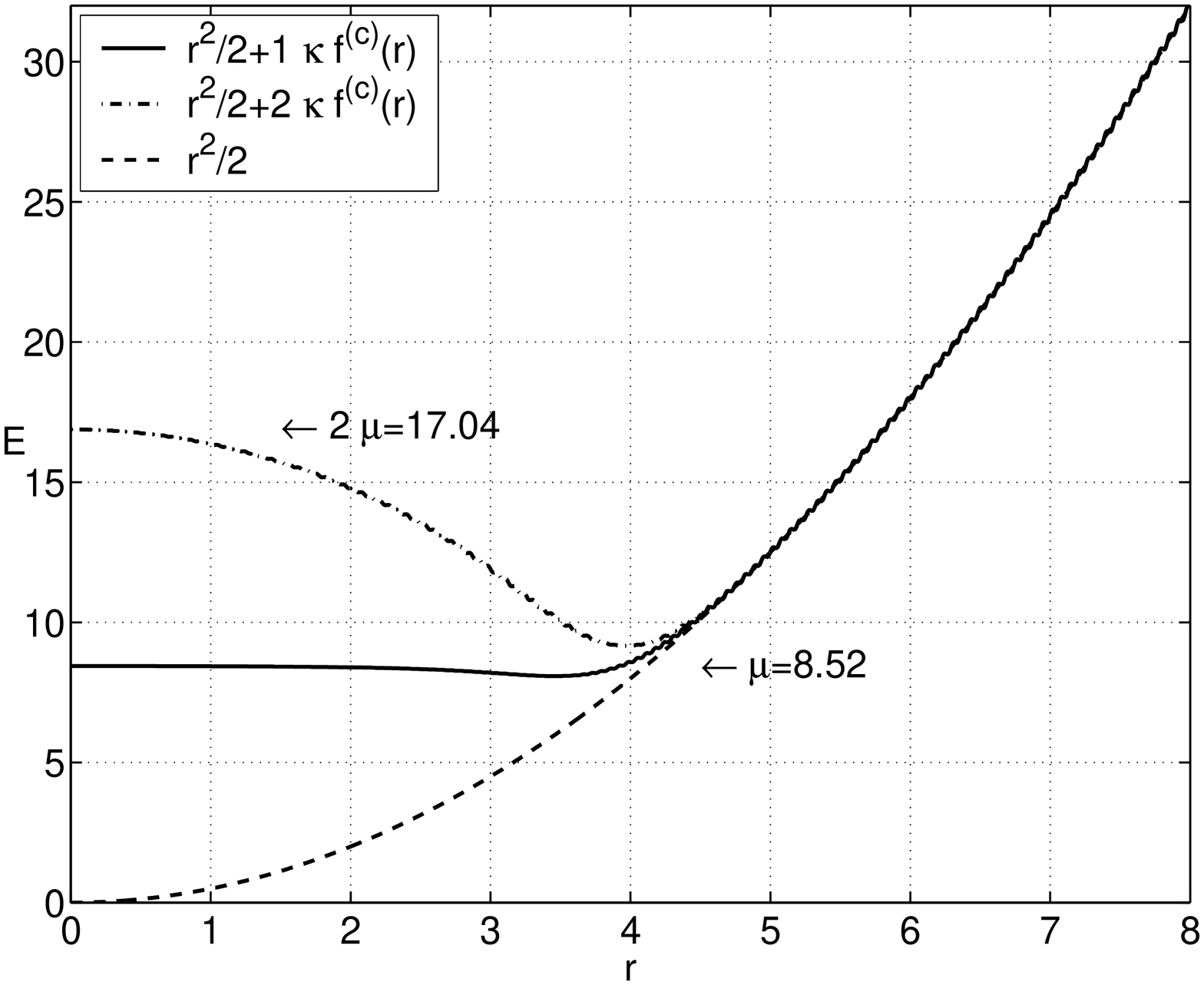,width=8.5cm,angle=90}
    \begin{caption}
      {\sf Self-consistent potential energy densities of the
        condensate Hamilton operator $\piGP{N}$ (solid), the
        noncondensate Hamilton operator $\sigN$ (dashed dot), and the
        bare harmonic oscillator potential (dashes) versus radius.
        Energy and length are scaled in the natural units for a
        harmonic oscillator.}
      \label{fig_pot}
    \end{caption}
\end{center}
\end{figure}

The position densities of the condensate $\fc{}(r)$, as well as the
local densities of the normal and anomalous fluctuations,
\ie, $\fs{}(r)$ and $\ms{}(r)$, are represented in
Fig.~\ref{fig_dens}.  First of all, it has to be noted that the
solutions are real valued.  This important fact follows from the
detailed structure of the stationary
Eqs.~(\ref{firstorderequilibriumc}) and
(\ref{firstorderequilibriumsq}), which are invariant under a global
phase change.  Second, if we focus on the condensate density, one can
see that it closely follows the Thomas-Fermi approximation
$\fc{\text{TF}}(r)=(r_{\text{TF}}^2-r^2)/(2\kappa)$, for radii less
than $r_{\text{TF}}=\sqrt{2\mu}\approx 4.12$.  This limit is valid in
the strong coupling regime $N^{(c)}\,\kappa >1$, where the kinetic
energy is a negligible contribution compared to the external trapping
potential and the self-energies.  The self-consistent solutions for
the normal and anomalous densities are depicted in the lower half of
Fig.~\ref{fig_dens}.  While all normal densities are necessarily
positive, the anomalous fluctuations carry a negative sign. The
anomalous fluctuations are the response of the noncondensate medium to
a phase-coherent mean-field.  In analogy to the polarization of an
atom that is subjected to an electric field, it tries to compensate
for the external perturbation.  In the evaluation of the normal and
anomalous fluctuations, we have truncated the finite temperature sums
beyond the radial and angular momentum quantum numbers $n^r=14$ and
$l=6$. This leads to fully converged values of the normal
fluctuations.  However, it has to be noted that the values of the
anomalous fluctuations are still subject to change (further decrease).
As long as one keeps adding vacuum contributions at an unaltered
strength of the s-wave matrix-element $\phi^{n_1^r l_1 n_2^r l_2 n_3^r
  l_3 n_4^r l_4}$ [see Eq.~(\ref{swavematel})], this would lead to the
well known ultra-violet divergence
\cite{Stoof1999a,huang,Hutchinson1998a}. Thus, a judicious truncation,
or alternatively an energy renormalization is needed to remove the
non-physical divergence that arises solely from the energy-independent
approximation of the scattering amplitudes.
\begin{figure}[f]
  \begin{center}
    \psfig{figure=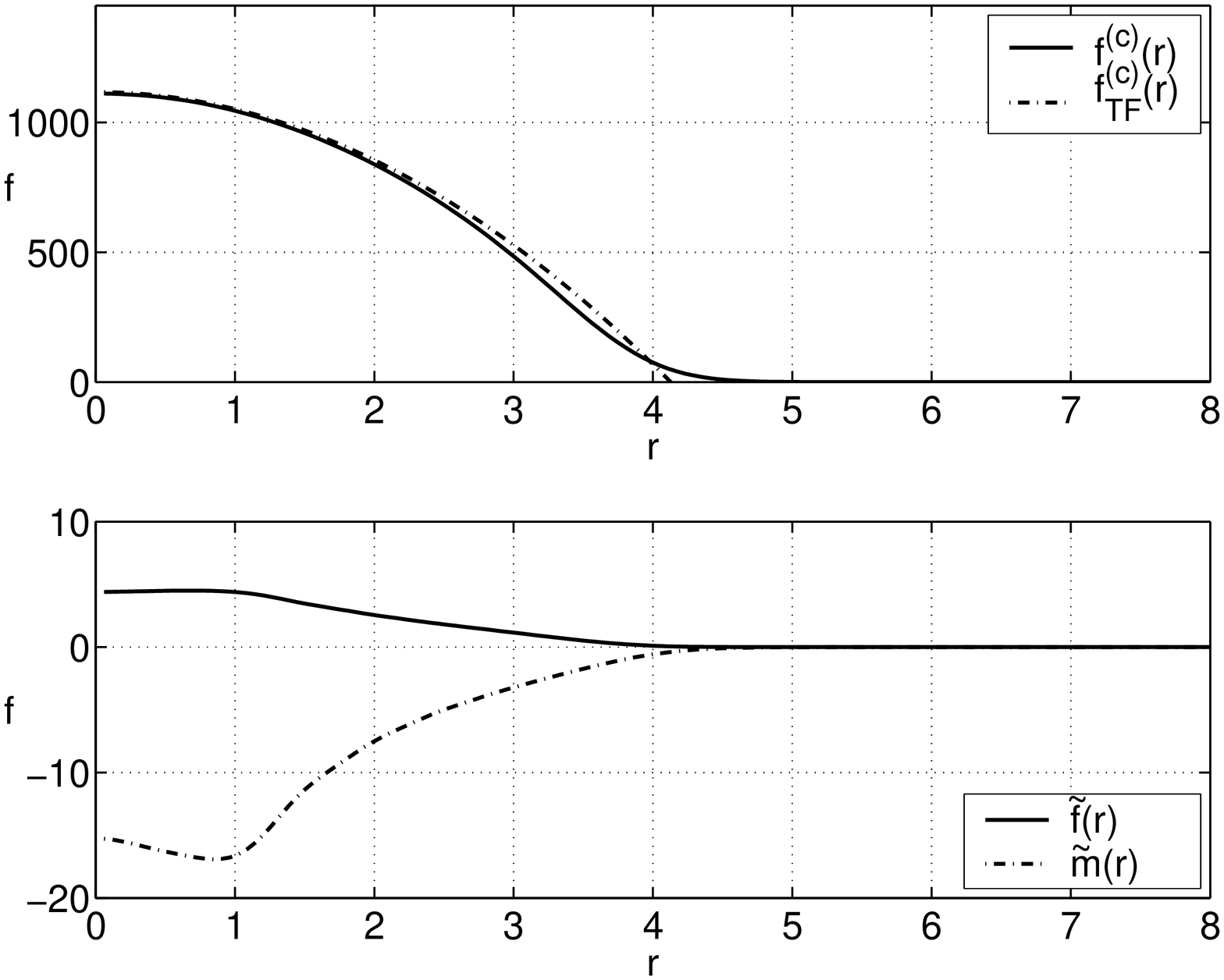,width=8.5cm,angle=90}
    \begin{caption}
      {\sf Position density of the mean-field $\fc{}(r)$ (solid), the
        Thomas-Fermi approximation $\fc{\text{TF}}(r)$ (dashed-dot),
        as well as the normal fluctuations $\fs{}(r)$ (solid) and the
        anomalous fluctuations $\ms{}(r)$ (dashed-dot). Density and
        length are scaled in the natural units for a harmonic
        oscillator.}
      \label{fig_dens}
    \end{caption}
\end{center}
\end{figure}

In Fig.~\ref{fig_particlestate}, we show a few selected radial
eigen-functions
$\scal{\bx}{\tilde{1}}=R^{n^r}_{l}(r)\,Y^{l}_{m}(\theta,\varphi)$ of
the noncondensate Hamiltonian operator
$(H^{(0)}+2\,\Uc)\,|\tilde{1}\rangle=\varepsilon_{1}\,|\tilde{1}\rangle$.
The lowest energy state $R^{n^r=1}_{l=0}(r)$ is localized at the rim
of the condensate $r_{\text{TF}}=\sqrt{2\, \mu} \approx 4.12$.  It has
a smaller spatial extent than the condensate and consequently a higher
energy.  All s-wave functions ($l=0$) have a finite value at the
origin in contrast to the $l>0$ states that must be vanishing at
$r=0$.  An eigen-function that is characterized by quantum numbers
$(n^r,l)$ has $n^r-1$ nodes.  Eigen-functions corresponding to higher
angular momenta $l$ are shifted outwards due to the increased angular
momentum barrier $l(l+1)/r^2$.  The corresponding eigen-energies are
depicted in Fig.~\ref{fig_HFB}.  In the context of spatially
homogeneous condensed matter systems, these eigen-functions are
associated with particle-like excitations.

\begin{figure}[ft]
  \begin{center}
    \psfig{figure=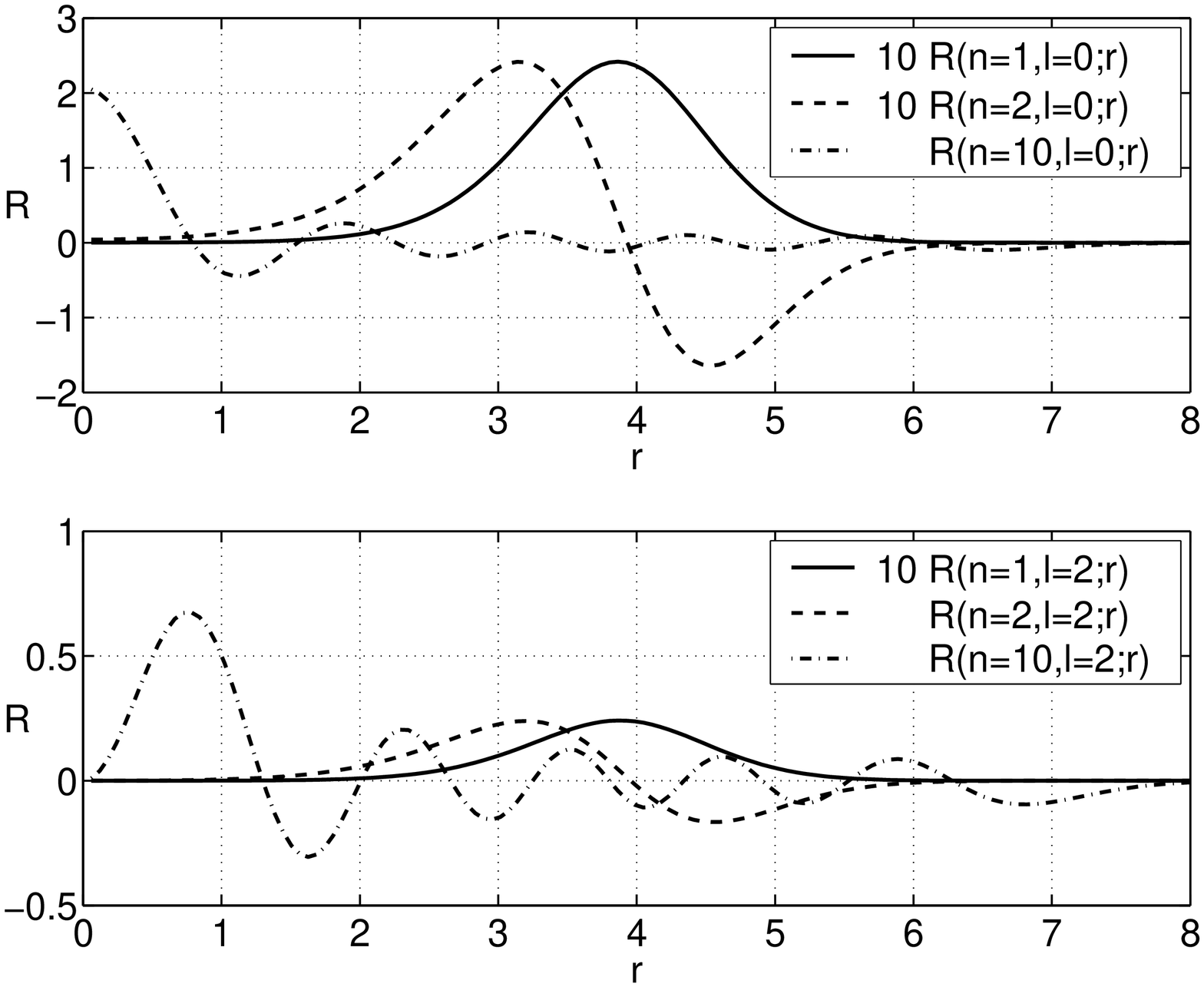,width=8.5cm,angle=90}
    \begin{caption}
      {\sf Radial eigen-functions $R^{n^r}_{l}(r)$ of the particle-like
        basis states $|\tilde{1}\rangle$ associated with the
        noncondensate Hamiltonian operator $H^{(0)}+2\,\Uc$ versus radius.
        Depicted are  a few representative states for the
        radial and angular quantum numbers $n^r=1,2,10$ and $l=0,2$.}
      \label{fig_particlestate}
    \end{caption}
\end{center}
\end{figure}
The other relevant set of eigen-states arises from the condensate
Hamiltonian operator, \ie, the stationary GP-equation
$(H^{(0)}+1\,\Uc)\,|1^{(c)}\rangle=\varepsilon_{1}\,|1^{(c)}\rangle$.
The lowest self-consistent energy eigen-state defines the condensate
wave function.  A selection of these eigen-states are shown in
Fig.~\ref{fig_phononstates}.  As these states correspond to the low
energetic excitation-modes of the condensate they are referred to as
phonon-like. The eigen-energies of the isotropic $(l=0)$ modes are
shown in Fig.~\ref{fig_HFB}.

We have compiled the four important positive energy spectra that arise
in the problem in Fig.~\ref{fig_HFB}. In essence, these are the
spectra of the condensate and the noncondensate Hamiltonians,
$\piGP{N}$ and $\sigN$, as well as their generalization in terms of
the HFB self-energy operators $\Pi$ and $\Sigma$.  It can be seen that
the s-wave energies of $\piGP{N}$ and $\Pi$ are virtually identical.
In contrast to this, one finds that the excitation frequencies of the
fluctuations $\Sigma$ are characteristically shifted downwards from
the energies of the noncondensate $\sigN$. It is also important to
note that the self-energy of the fluctuations $\sigN$ includes the
energy shifts of the noncondensate itself.  These numerical results
compare well within the limits of validity with the perturbative and
semi-classical approximations of
Refs.~\cite{Stringari1996a,Minguzzi1997a,Csordas1998a}.

The spectrum of eigen-values of $\Sigma$ exhibits a characteristic
energy gap above the condensate energy level $\mu$. It is well known
that  this gap energy vanishes asymptotically for a homogeneous
system in the thermodynamic limit.  By
deliberately excluding the anomalous coupling $\Vsq$ strength from a
first order theory (Popov approximation) or by including the second
order Beliaev correction of Eq.~(\ref{Boltzeq}) in the self-energies,
one can obtain a gapless approximation  \cite{Hohenberg1965}.
\begin{figure}[ft]
  \begin{center}
    \psfig{figure=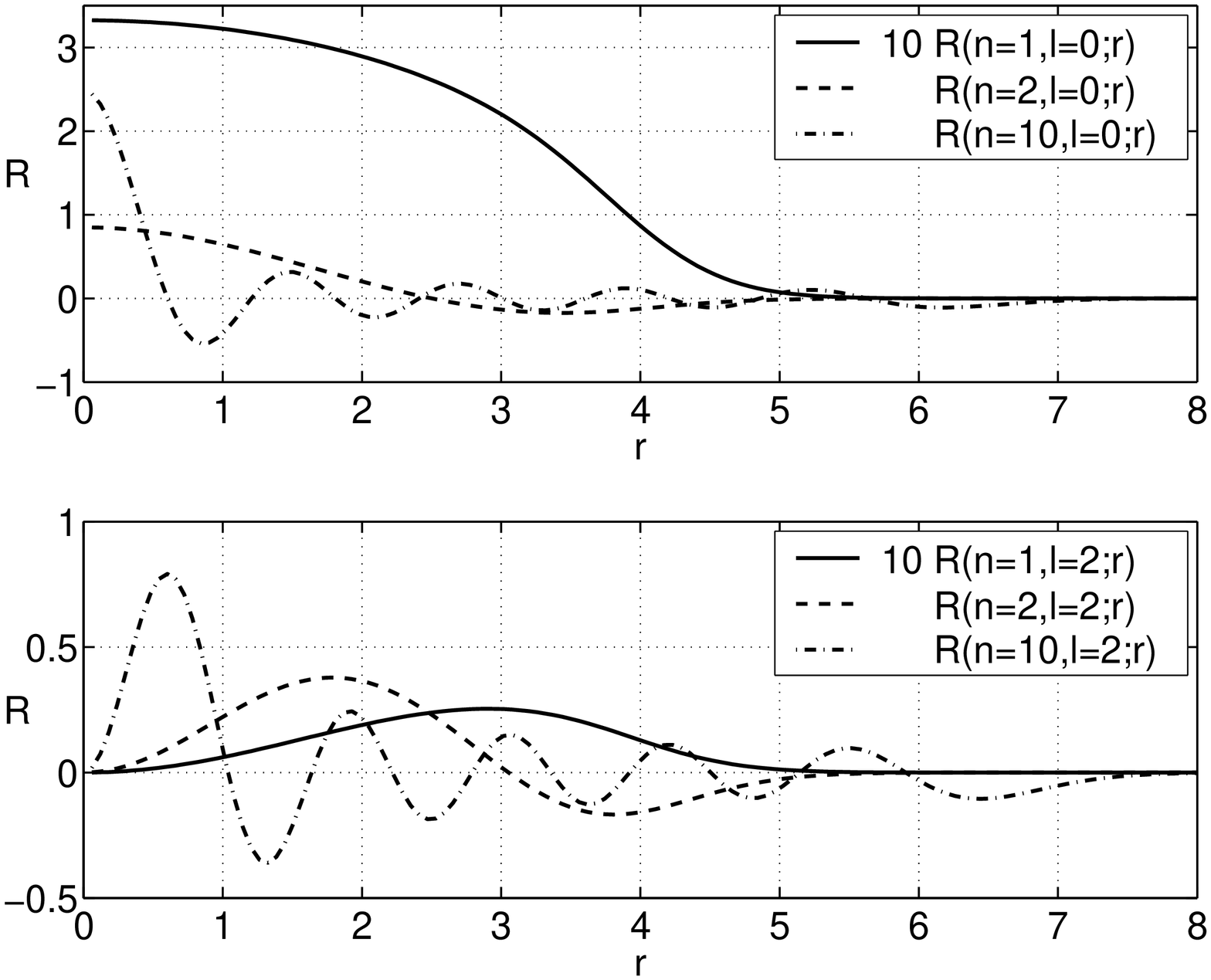,width=8.5cm,angle=90}
    \begin{caption}
      {\sf Radial eigen-functions $R^{{n^r}}_l(r)$ of the phonon-like
        basis states $|1^{(c)}\rangle$ of the mean-field Hamiltonian
          operator $H^{(0)}+1\,\Uc$ versus radius. Depicted are a few
          representative states for the radial and angular quantum
          numbers $n^r=1,2,10$ and $l=0,2$.}
      \label{fig_phononstates}
    \end{caption}
\end{center}
\end{figure}
\subsection{Time-dependent solution of the Hartree-Fock-Bogoliubov equations}
After studying some aspects of the stationary solutions of the
generalized HFB equations Eqs.~(\ref{firstorderequilibriumc}) and
(\ref{firstorderequilibriumsq}), such as the local densities, the
eigen-states, or the energy spectra, we will now investigate the
reversible real time evolution of the coupled condensate and
noncondensate system, \ie, \bea
\label{GPeqrev}
\frac{d}{dt} \chi
&=&-i\, \Pi\, \chi,\\
\label{Boltzeqrev}
\frac{d}{dt} \Gg &=&-i\Sigma\,\Gg
+i\,\Gg\,\Sigma^\dag.  \eea In contrast to the complete
kinetic Eqs.~(\ref{GPeq}) and (\ref{Boltzeq}), which account for
collisionally induced energy shifts and irreversible population
transfer, Eqs.~(\ref{GPeqrev}) and (\ref{Boltzeqrev}) contain only the
first order reversible processes. As we will show in the following, this
restriction implies constant occupation numbers $P$.
\begin{figure}[ht]
  \begin{center}
    \psfig{figure=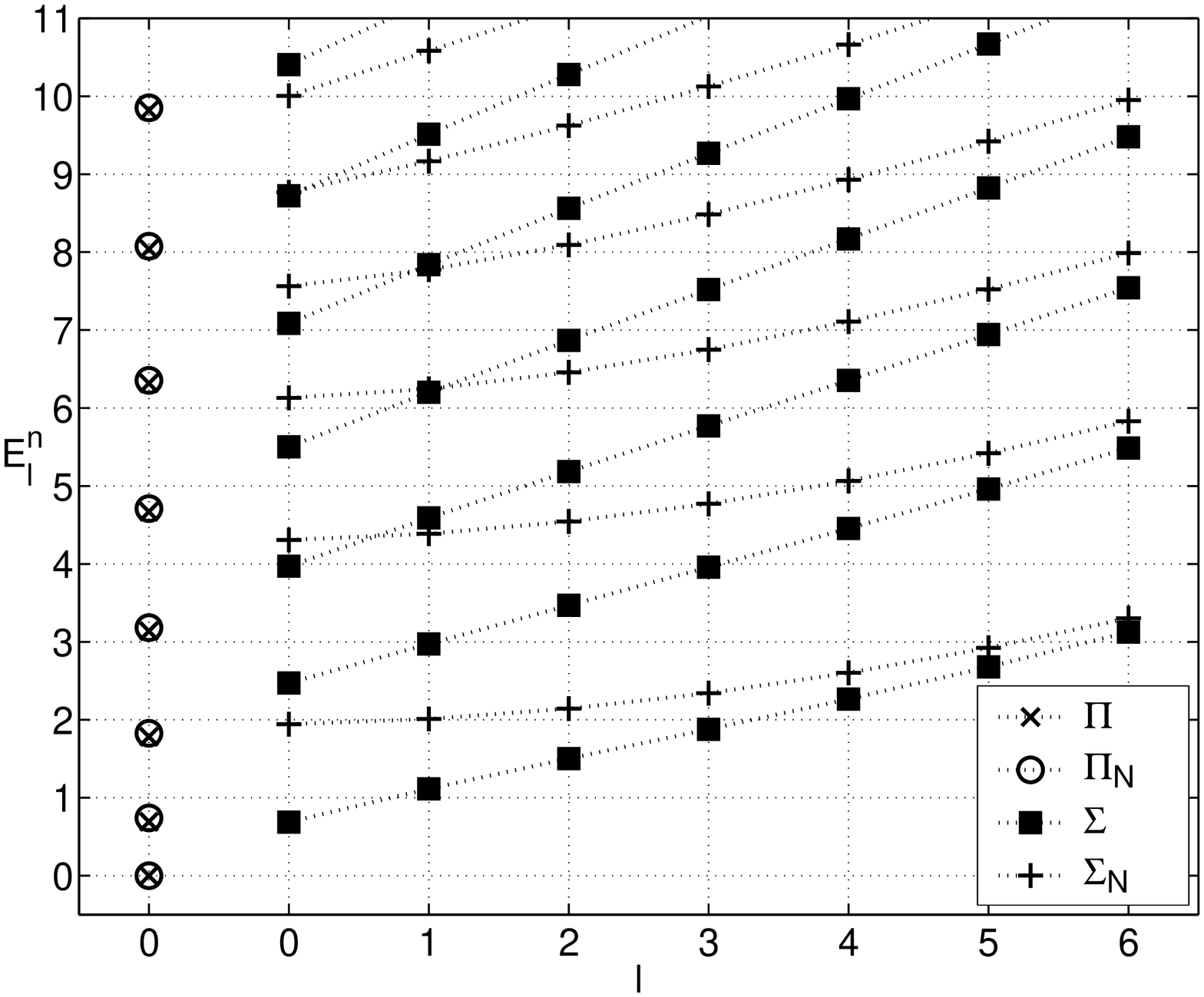,width=8.5cm,angle=90}
    \begin{caption}
      {\sf Energy spectra $E_{l}^{n^r}$ versus radial and angular
        quantum numbers $1\leq n^r \leq 8$ and $0 \leq l\leq 6$ for
        the phonon-like states of $\piGP{N}$ ($l=0$, only), the
        particle-like states of $\sigN$, as well as the positive part
        of the energy spectrum of $\Pi$ ($l=0$, only) and HFB
        self-energy $\Sigma$. Dimensionless energies are measured in
        natural harmonic oscillator units.}
      \label{fig_HFB}
    \end{caption}
\end{center}
\end{figure}
In Sec.~\ref{secG}, we have shown that any admissible fluctuation
matrix $\Gg$ has to be of the form \bea \Gg&=&V_{+}^{\phantom{\dag}}\,
P \, V_{+}^\dag +V_{-}^{\phantom{\dag}}\,(1+P)\,V_{-}^\dag, \eea where
P represents the positive occupation numbers of the eigen-states
$V_{+}$.  This property is not only to be satisfied in equilibrium
where the eigen-states coincide with the HFB states [see
Eq.(\ref{finiteTG})], but in all instances.

By formally integrating the reversible kinetic equations, we can show
that this structure of the fluctuation matrix is preserved at all
times. Thus, Eqs.~(\ref{GPeqrev}) and (\ref{Boltzeqrev}) define a
consistent initial value problem.  This simple but important fact can
be demonstrated easily by defining the formal solution in terms of a
time-ordered exponential: \bea T(t,t_0)&=& {\mathcal{T}}
\exp{[-i\int_{t_0}^{t}dt^\prime\,\Sigma(t^\prime)]}.  \eea It is
obvious that the propagator matrix $T(t,t_0)$ can be constructed in
two steps by $T(t,t_0)=T(t,t_1)\,T(t_1,t_0)$ since the self-energy is
local in time (semi-group property).  Moreover, it follows from the
structure of the generator that $T(t,t_0)$ is a proper symplectic
transformation $\paul{3}=T(t,t_0)\,\paul{3}\,T(t,t_0)^\dag$ at all
times.  Consequently, all occupation numbers of the general solution
to the nonlinear, initial value problem are constants of motion \bea
\label{instantG}
\Gg(t)&=&T(t,t_0)\,\Gg(t_0)\, T(t,t_0)^\dag\\
&=& V_{+}^{\phantom{\dag}}(t)\, P(t_0) \,V_{+}^\dag(t)
+V_{-}^{\phantom{\dag}}(t)\,(1+P(t_0))\,V_{-}^\dag(t), \nonumber
\eea with time-evolved basis
states $V(t)=T(t,t_0)\, V$. Since $T(t,t_0)$ represents a genuine
symplectic transform, the eigen-states
$V^\dag(t)\,\paul{3}\,V(t)^{\phantom{\dag}}=
V^\dag\,\paul{3}\,V^{\phantom{\dag}}=\paul{3}$ remain orthogonal.

In the following figures, we illustrate these fundamental facts that
effectively define reversibility for any non-linear system in the
generic form of Eqs.~(\ref{GPeqrev}) and (\ref{Boltzeqrev}).  In
particular, we will use the self-consistent, finite temperature
solution for $\chi(t=0)=\chi(\beta)$ and $\Gg(t=0)=\Gg(\beta)$
as an initial value for the time-propagation at $t=0$. It is obvious
that this choice does not induce any change during the subsequent
real-time evolution.  However, at $t=1$, we suddenly distort this
equilibrium solution by setting the anomalous component of $\Gg(t=1_-)$
to zero, \ie, $\ms{}(t=1_+)=0$ and then propagate forward up to
$t=4$.  It is important to note that the new $\Gg(t=1_+)$ is still a
valid fluctuation matrix.
\begin{figure}[ht]
  \begin{center}
    \psfig{figure=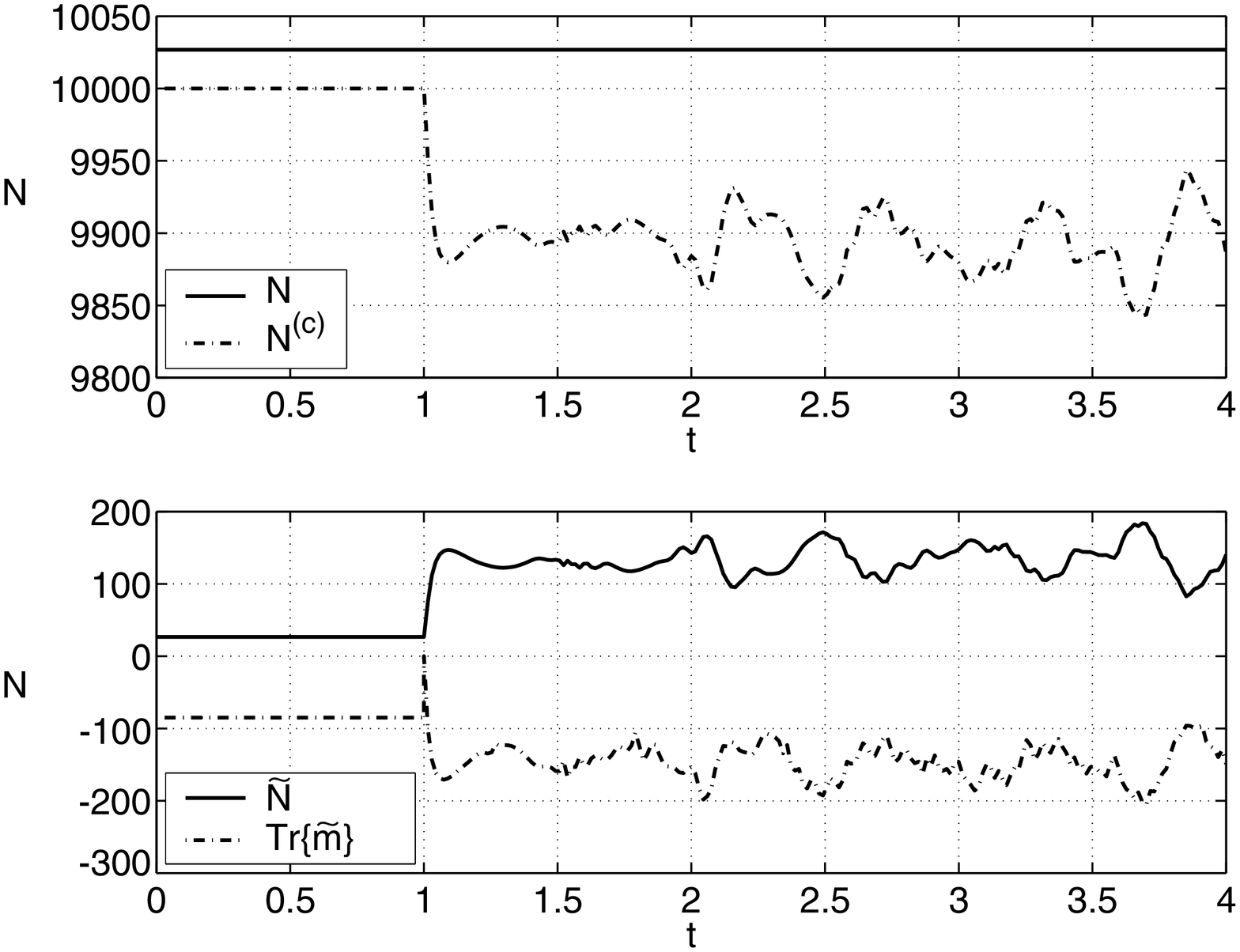,width=9cm,angle=90}
    \begin{caption}
      {\sf Real time evolution of the total particle number
        $N=N^{c}+\widetilde{N}$ (solid), the number of particles in
        the condensate $N^{c}$ (dash-dot), the noncondensate particles
        number $\widetilde{N}$ (solid) and the trace over the
        anomalous fluctuations $\text{Tr}\{ \ms{}\}$ (dashed-dot).  At
        $t=1$, the equilibrium solution is suddenly distorted by
        setting $\ms{}(t=1)=0$. Dimensionless time is measured in
        natural harmonic oscillator periods.}
      \label{fig_number}
    \end{caption}
\end{center}
\end{figure}
In Fig.~\ref{fig_number}, we have depicted the number of particles that
occupy the condensate $N^{(c)}=\text{Tr}\{\fc{}(t)\}$, the
noncondensate $\tilde{N}=\text{Tr}\{\fs{}(t)\}$, and the total
particle number $N=N^{(c)}+\tilde{N}$ versus time. Time is measured in
units of the harmonic oscillator period $T=2\pi/\omega$. In contrast
to these numbers that are genuine single-particle properties, the
anomalous fluctuations are a physical measure of the degree of
two-particle correlations or squeezing \cite{Naraschewski996}. For
example, the total particle number fluctuations
$\av{(\hat{N}-\av{\hat{N}})^2}$ or, more specifically, the normally
ordered density fluctuations
$\av{:\hat{f}(\bx,\bx),\hat{f}(\by,\by):}$ would reflect the degree of
squeezing of the quadrature components along certain directions $\bx$,
$\by$. While we have not evaluated such observables here, we have included
the averaged strength of the anomalous fluctuations
$\text{Tr}\{\ms{}\}$ to represent their size.

The most important feature in Fig.~\ref{fig_number} is the exact
conservation of the total particle number during all phases of the
evolution.  The instantaneous change in $\ms{}(t=1)$ does not affect it
directly. But it can be seen that the relative partitioning of the
particles between condensate and normal noncondensate is massively
distorted by this sudden influx of energy and particles start to oscillate
coherently between the coupled subsystems.

In Fig.~\ref{fig_eng}, we show the individual first order
contributions to the total system energy $E=E^{(c)}+E[\fs{}]+E[\ms{}]$
as described in Eqs.~(\ref{Ec})--(\ref{Ems}). Again the most notable
feature is the exact conservation of total energy during the time-evolution. 
Due to the sudden change in the anomalous fluctuations at $t=1$, the overall
energy increases instantaneously by more than $400\, \hbar \omega$. 
\begin{figure}[ht]
  \begin{center}
    \psfig{figure=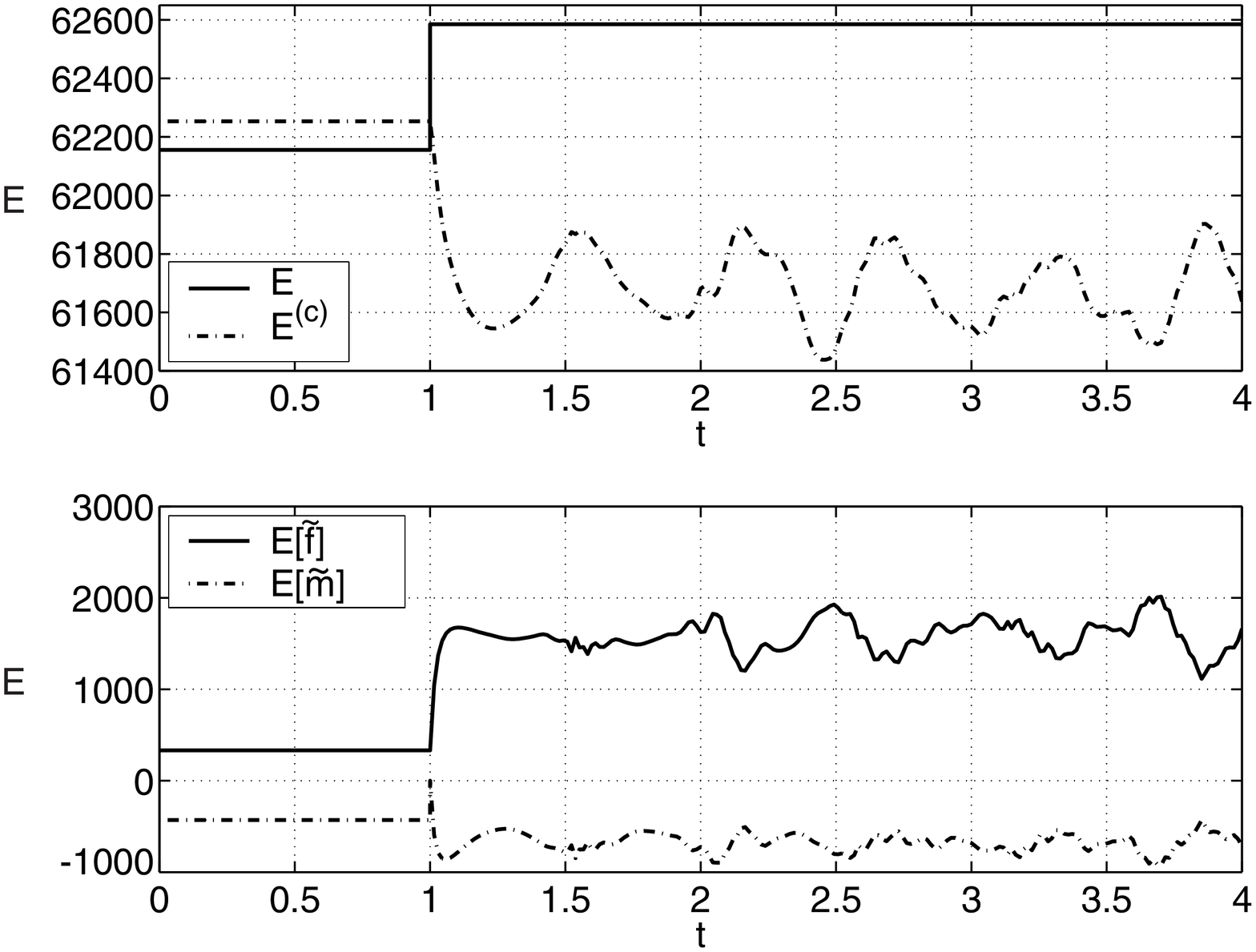,width=9cm,angle=90}
    \begin{caption}
      {\sf Real time evolution of the total system energy
        $E=E^{(c)}+E[\fs{}]+E[\ms{}]$ (solid), the energy of the
        particles in the condensate $E^{(c)}$ (dashed-dot), the normal
        noncondensate energy $E[\fs{}]$ and the anomalous energy
        $E[\ms{}]$. At $t=1$, the equilibrium solution is suddenly
        distorted by setting $\ms{}(t=1)=0$.  Dimensionless energy is
        measured in natural harmonic oscillator units.}
      \label{fig_eng}
    \end{caption}
\end{center}
\end{figure}
In Fig.~\ref{fig_occup}, we show the complete spectrum of occupation
numbers $P(1\leq n^r \leq 14,0\leq l \leq 6;t)$ that occur in the
instantaneous fluctuation matrix $\Gg(t)$, as defined by
Eq.~(\ref{instantG}).  From the logarithmic plot that covers 13
decades, it can be seen that the occupation numbers are indeed
numerically exact constants of motion. Due to the instantaneous change
at $t=1$, many of the high-energy occupation numbers that are
quasi-degenerate before the change, split up into a multitude of
nondegenerate levels afterwards.  Minute changes in the occupation
numbers $|\delta P| <10^{-10}$ at the end of the integration period
identify the precision loss of the numerical calculation.
\begin{figure}[ht]
  \begin{center}
    \hspace{-10mm} \vspace{-10mm}

    \psfig{figure=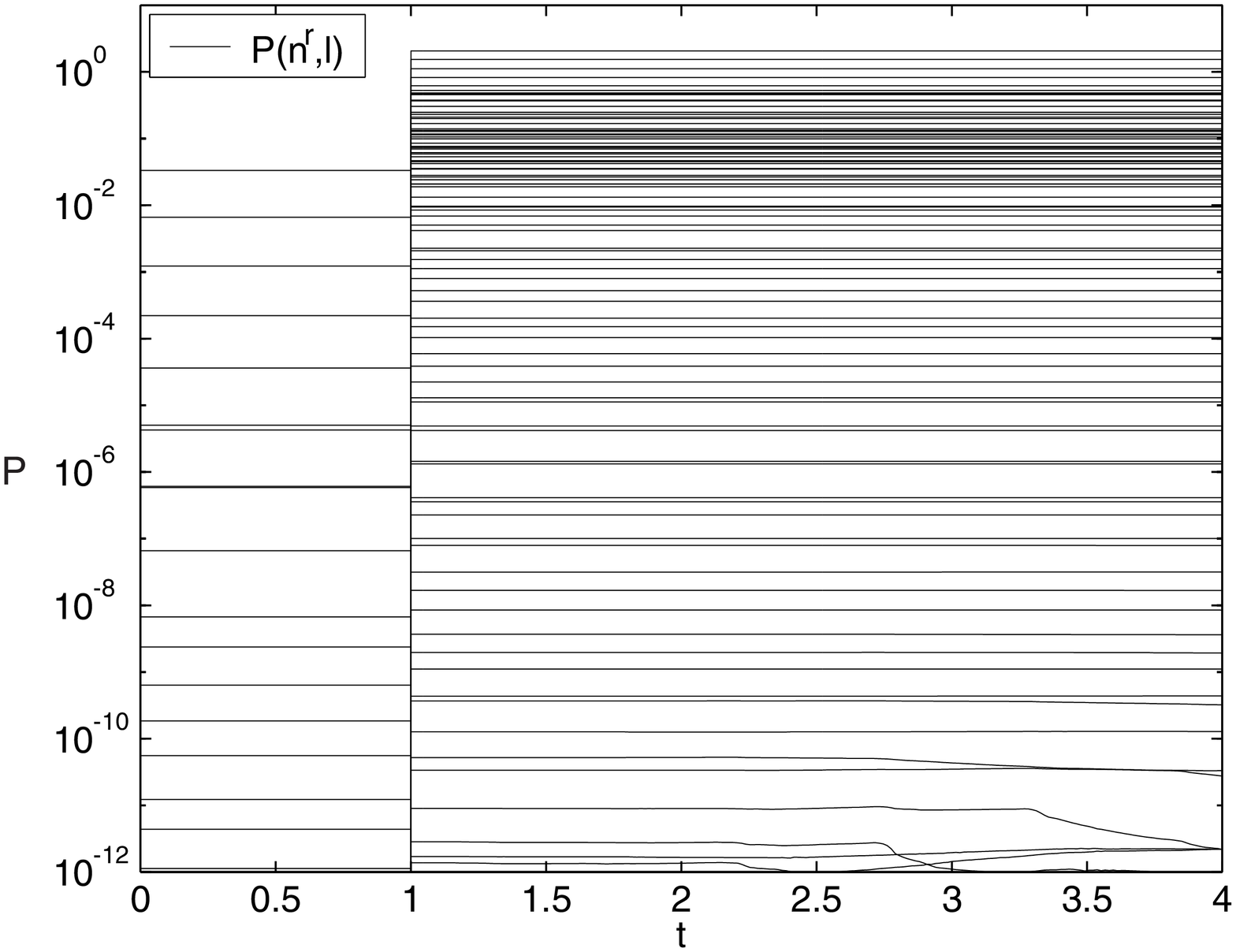,width=9cm,angle=90}
    \begin{caption}
      {\sf Real time evolution of the instantaneous occupation numbers
        $P(1\leq n^r\leq 14,0 \leq l \leq 6)$ that characterize the
        fluctuation matrix $\Gg(t)$. At $t=1$,
        $\Gg(t)$ is suddenly distorted by setting $\ms{}(t=1)=0$.}
      \label{fig_occup}
    \end{caption}
\end{center}
\end{figure}
The numerical results discussed in this section were obtained by
discretizing the generalized GP Eq.~(\ref{GPeqrev}) and the reversible
part of the Boltzmann Eq.~(\ref{Boltzeqrev}) with a standard finite
element method \cite{EsryThesis97} based on b-splines
\cite{deboor78a,pppack}.  The use of these finite-support, piecewise
polynomial basis functions results in matrix representations of the
kinetic and potential energy operators that are banded. Very efficient
linear algebra algorithms can be employed in this case (LAPACK
\cite{lapack}).  In particular, we used a set of 400 b-splines on an
equidistant radial grid from $0\le r\le 25 \approx 6\, r_{\text{TF}}$.
Eventually, we represented the condensate wave-function $\al{}$
(Fig.~\ref{fig_dens}), the particle-like basis states
$|\tilde{1}\rangle$ (Fig.~\ref{fig_particlestate}), as well as the
phonon-like basis sates $|1^{(c)}\rangle$ (Fig.~\ref{fig_phononstates})
in this b-spline basis.  Finally, a subset of quantum states was
chosen (either $|\tilde{1}\rangle$ or $|1^{(c)}\rangle$) with
$\{(n^r\ge1,l\ge0): 2(n^r-1)+l<=18\approx 2\mu\}$ to evaluate the
finite temperature sums or to propagate in real time.  We have
verified numerically that no particular advantage can be obtained from
either choice, as long as all of the relevant energies scales are
resolved. The conservation of the total particle number $\av{N}$
(Fig.~\ref{fig_number}), the total energy $E$ (Fig.~\ref{fig_eng}) as
well as the instantaneous occupation numbers $P$
(Fig.~\ref{fig_occup}) support this argument.
\section{The irreversible evolution}
\label{irreversible}
\subsection{An ergodic equilibrium solution of the master-equation}
In the following discussion of the irreversible evolution of the
kinetic equations, we will again try to elucidate the main physics by
additional simplifications. In particular, we will assume ergodicity
for the normal fluctuations $\fs{12}=\delta_{12}\,\fs{\eps{1}}$, and
the anomalous fluctuations $\ms{}=0$.  This physical approximation is
appropriate for most kinetic temperatures, except for a region close
to $T=0$. Within this limit, we are able to establish an important
result for the stationary behavior of the condensed atomic gas, \ie,
a canonical Bose-Einstein distribution for the noncondensate particles
coexisting with an energetically lower-lying, coherent condensate
mode.  By imposing the restriction of vanishing anomalous
fluctuations upon the kinetic equations, [Eq.~(\ref{GPeq}) and
(\ref{Boltzeq})], we are left with the following equations of motion:
\bea
\label{GPsimple}
\frac{d}{dt} \al{}&=&(-i\, \piGP{N}+\Upk{N}-\Upg{N})\,\al{},\\
\label{Boltzeqsimple}
\frac{d}{dt}
\fs{}&=&\Gak{N}\,\fspo{}-\Gag{N}\fs{}+\text{h.{\thinspace}c.}\eea It
is worth mentioning that in the real time evolution the rotating frame
frequency $\mu$ is still an adjustable parameter and not necessarily
synonymous with the chemical potential.

First, let us concentrate on the equation for the mean-field amplitude
$\al{}$. From Eq.~(\ref{GPsimple}), we can see that the mean-field
evolution consists of two distinct parts: a Hermitian,
number-conserving contribution $\piGP{N}$, and a part that accounts
for condensate number changing collisions out of and into the
noncondensate, \ie,  $\Upk{N}-\Upg{N}$.  In stationarity, both processes have
to vanish identically
\bea
\label{eqfirst}
0&=&(H^{(0)}+1\,\Uc+2\,\Usq-\mu)\,\al{},\\
\label{eqlater}
0&=&( \Gcoll{\fs{} \fs{} \fspo{}}-\Gcoll{\fspo{} \fspo{}
  \fs{}}\,)\,\al{}.  \eea Given the constraint on the condensate
particle number, it is in principle straightforward to solve the
Hermitian eigen-value problem of Eq.~(\ref{eqfirst}) and obtain an
eigen-value $\mu$. The later equation Eq.~(\ref{eqlater}) poses a much
more challenging constraint on the coupled system. In order to
maintain a stationary state, the mean-field has to be ``orthogonal'' to
all number changing processes.  This means that the normal
fluctuations have to adjust self-consistently with respect to the
condensate wave-function.  It is interesting to note that the
solutions of this system are infinitely degenerate with respect to a
global phase-rotation. In other words, the condensate's phase is not
pinned down by any restoring force and is free to drift, consequently.
However, the later, vector-valued condition of Eq.~(\ref{eqlater}) is
satisfied identically if \bea
\label{balancecond}
0&=&\phi^{4^\ppr 1\, 2^\ppr 1^\ppr} \phippr{}\,
\delta_{\eta}(\epspp{1}+\epspp{2}-\mu-\epspp{4})\nonumber\\
&&(1+\fs{\epspp{1}})\, (1+\fs{\epspp{2}})\,
\alpp{3}\,(1+\fs{\epspp{4}})\nonumber\\
&& [ \frac{ \fs{\epspp{1}} }{ (1+\fs{\epspp{1}}) } \frac{
  \fs{\epspp{2}} }{ (1+\fs{\epspp{2}}) }- \frac{ \fs{\epspp{4}} }{
  (1+\fs{\epspp{4}}) } ], \eea vanishes for all components $1$.
Provided that energy conservation is satisfied exactly, \ie, 
$\epspp{1}+\epspp{2}=\mu+\epspp{4}$, it is straightforward to verify
that a canonical Bose-Einstein distribution \bea
\label{BEdist}
\fs{\eps{}}&=&\frac{1}{e^{\,\beta(\eps{}-\mu)}-1}, \eea with an
inverse temperature $\beta$, is the equilibrium solution.  In addition
to this functional form of the distribution that is dictated by
detailed balance, it is required that all of the excitation energies
are above the condensate energy, \ie, $\eps{1}> \mu$. Thus the
eigen-energy spectrum exhibits a finite gap.  From Fig.~\ref{fig_HFB}
it can be seen that both the positive energy spectrum of the HFB
operator $\Sigma$ as well as the eigen-energies of $\sigN$, are
suitable candidates within this approximation.

Second, the generalized Boltzmann equation, Eq.~(\ref{Boltzeqsimple}),
is stationary if \bea
\label{boltzii}
0&=&( \Gcoll{\fs{} \fs{} \fspo{}}+ 2\,\Gcoll{\fc{} \fs{} \fspo{}}
+\Gcoll{\fs{} \fs{} \fc{}})
\fspo{}\\
&-&( \Gcoll{\fspo{} \fspo{} \fs{}} + 2\,\Gcoll{\fc{} \fspo{} \fs{}}
+\Gcoll{\fspo{} \fspo{} \fc{}})\fs{}.\nonumber \eea Within the ergodic
approximation, the Hermitian part is satisfied identically. On the
other hand, there are two distinct types of collisional relaxation
processes in Eq.~(\ref{boltzii}). There are the number-conserving in
and out rates of the conventional quantum-Boltzmann equation, \ie, $
\Gcoll{\fs{} \fs{} \fspo{}}\fspo{}- \Gcoll{\fspo{}
  \fspo{}\fs{}}\fs{}$.  Obviously, both rates match identically under
the detailed balance conditions of Eq.~(\ref{BEdist})
\bea
0&=&\phi^{4^\ppr 1\, 2^\ppr 1^\ppr} \phi^{1^\ppr 2^\ppr 2\, 4^\ppr}\,
\delta_{\eta}(\epspp{1}+\epspp{2}-\epspp{4}-\eps{2})\nonumber\\
&&(1+\fs{\epspp{1}})\,(1+\fs{\epspp{2}})\,(1+\fs{\epspp{4}})\,
(1+\fs{\eps{2}})\nonumber\\
&& [ \frac{ \fs{\epspp{1}} }{ (1+\fs{\epspp{1}}) } \frac{
  \fs{\epspp{2}} }{ (1+\fs{\epspp{2}}) } - \frac{ \fs{\epspp{4}}
  }{(1+\fs{\epspp{4}}) } \frac{ \fs{\eps{2}} }{(1+\fs{\eps{2}}) }
].\eea
Both of the two remaining distinct processes in Eq.~(\ref{boltzii}),
\ie, $2\,\Gcoll{\fc{}\fs{}\fspo{}}\,\fspo{} -2\,\Gcoll{\fc{} \fspo{}
  \fs{}}\,\fs{}$, as well as the process $\Gcoll{\fs{} \fs{}
  \fc{}}\,\fspo{} -\Gcoll{\fspo{} \fspo{} \fc{}}\,\fs{}$, involve a
condensate particle in the pre-collision or post-collision channels.
Thus, real particles will be transfered between the condensate and
the noncondensate, until the rates are balanced.  Analogous to the
arguments that lead to Eq.~(\ref{balancecond}), it can be shown that a
canonical \BE distribution is attained in equilibrium.

\section{Conclusions and Outlook}

In this article, we have studied aspects of the reversible and
irreversible evolution of a condensed atomic gas immersed in the
noncondensate. By specializing the kinetic equations
\cite{Walser1999a} for a simple isotropic model, we were able to
analyze the equilibrium solution, as well as the dynamic
nonequilibrium behavior numerically. In particular, we obtained the
excitation spectra of a finite temperature equilibrium. Moreover, we
demonstrated the reversibility of the time-dependent HFB equations far
from equilibrium and in the collisionless regime. This is tantamount
to noting that the instantaneous occupation numbers of the HFB modes
are constants of motion. Finally, we studied the collisional regime
for an ergodic distribution of quasi-particles and showed that
detailed balance is obtained in the full quantum kinetic theory with a
self-consistent canonical \BE distribution.

Based on this isotropic model, we can also obtain the collision rates
that lead to a self-consistent equilibrium. However, such an analysis
is still work in progress and results will be presented in future
publications.

\section*{Acknowledgments} 
R.~W. acknowledges gratefully financial support by the U.S. Department
of Energy and the Austrian Academy of Sciences through an APART grant.
This work also benefited greatly by the BEC seminars of
Professor~C.~Wieman, Professor~E.~Cornell, and Professor~D.~Jin, as
well as discussions with J.~Wachter.

\appendix
\section{Canonical transformations}
\label{canonical}   
A canonical transformation is an inhomogeneous linear combination of
creation and destruction operators that preserves the commutation
relation \cite{blaizot}.  In particular, if $\aop{}$ and
$\aopd{}$ denotes a pair of Hermitian conjugated bosonic operators,
such that
\bea [ \aop{1},\aopd{2} ]&=&\delta_{1,2}, \eea 
then any affine linear transformation defines a new set of 
operators $b$ and 
$\bar{b}$ by 
\bea
\label{affine}
\left(
  \begin{array}{c}
     b\\
    \bar{b}
  \end{array}
\right)
&=& T\,\left(
  \begin{array}{c}
    \aop{}\\
    \aopd{}
  \end{array}
\right)+d.  \eea In an $n$-dimensional vector space, $T$ represents a
$(2\,n\times2\,n)$ dimensional matrix and $d$ is a $(2\,n)$ dimensional
vector.  Such a transformation is canonical if the new pair of operators
also satisfies the commutation relation:  \bea
\label{commutb}
[ b_{1},\bar{b}_{2}]=\delta_{1,2}.  \eea More specifically, the
transformation is unitary canonical if the new operators are Hermitian
conjugate pairs, \ie, $\bar{b}=b^\dag$.  By inserting
Eq.~(\ref{affine}) into Eq.~(\ref{commutb}), one finds that the
transformation matrices are a representation of the symplectic group
$Sp(2\,n)$: \bea T\,\paul{3}\, T^\dag=\paul{3}. \eea In addition, it
can be shown that $T^{\ast}=\paul{1}\,T\,\paul{1}$ and
$T^{-1}=\paul{3}\,T^{\dag}\paul{3}$.  Here, we have introduced the
$(2\,n)$-dimensional Pauli matrices $\paul{1}$ and $\paul{3}$ as \bea
\begin{array}{cc}
\paul{1}=
\left(
  \begin{array}{cc}
    0 & 1\\
    1 & 0
  \end{array}
\right), &
\paul{3}=
\left(
  \begin{array}{cc}
    1 & 0\\
    0 & -1
  \end{array}
\right).
\end{array}
\eea

\section{Cauchy-Schwartz inequality}
\label{CauchySchwartz}
For  a positive semi-definite density operator $\sigma$
and a general operator $\hat{L}$ it 
follows that the expectation value
\bea
\label{genposexp}
\av{\hat{L}\, \hat{L}^\dag}&=& \text{Tr}\{\sigma\,\hat{L}\,
\hat{L}^\dag\}\ge 0 \eea is never negative.  Consequently, the
co-variance matrix $\Gg$ of Eq.~(\ref{generaldensity}) must be positive
semi-definite $u^\dag \Gg u \ge 0$, as well.  This can be easily seen,
by considering a linear combination of two arbitrary operators
$\hat{A}$ and $\hat{B}$, \ie, $L=\alpha \hat{A}+\beta
\hat{B}$.  By minimizing the positive expression
Eq.~(\ref{genposexp}), one obtains the Cauchy-Schwartz inequality as
\bea \av{\hat{A}\hat{A}^\dag}\av{\hat{B}\hat{B}^\dag}\ge
\av{\hat{B}\hat{A}^\dag}\av{\hat{A}\hat{B}^\dag}.  \eea In particular,
for the special choice of $\hat{A}=\aop{1}-\al{1}$ and
$\hat{B}=\aopd{2}-\ald{2}$, this implies that the magnitude of the
anomalous fluctuations is limited by \bea (1+\fs{11})\fs{22}\ge
|\ms{12}|^2.  \eea


\end{document}